\documentclass{article} 
\usepackage[utf8]{inputenc}
\usepackage{latexsym,amsmath,amsfonts,amssymb,bm,empheq,float}   
\usepackage{graphicx}
\usepackage{booktabs}
\usepackage{supertabular}
\usepackage[dvipsnames]{xcolor}
\usepackage{caption,subcaption}
\usepackage{geometry}
\usepackage{hyperref}
\usepackage[normalem]{ulem}
\usepackage{authblk}
\usepackage{lipsum}

\begin{document}
\title{The big bang of an epidemic}

\author[1]{Yazdan Babazadeh}
\author[2]{Amin Safaeesirat}
\author[3,*]{Fakhteh Ghanbarnejad} 
\affil[1]{\small{Department of Applied Mathematics, University of Waterloo, Waterloo, ON N2L 3G1, Canada}}
\affil[2]{\small{Department of Physics, Simon Fraser University, Burnaby, Canada}}
\affil[3]{\small{Potsdam Institute for Climate Impact Research (PIK), Member of the Leibniz Association, P.O. Box 601203, Potsdam, 14412, Germany}}
\affil[*]{fakhteh.ghanbarnejad@gmail.com}


\maketitle

\begin{abstract}
In this paper, we propose a mathematical framework that governs the evolution of epidemic dynamics, encompassing both intra-population dynamics and inter-population mobility within a metapopulation network. By linearizing this dynamical system, we can identify the spatial starting point(s), namely the source(s) (A) and the initiation time (B) of any epidemic, which we refer to as the "Big Bang" of the epidemic. Furthermore, we introduce a novel concept of effective distance to track disease spread within the network. Our analysis reveals that the contagion geometry can be represented as a line with a universal slope, independent of disease type (R0) or mobility network configuration. The mathematical derivations presented in this framework are corroborated by empirical data, including observations from the COVID-19 pandemic in Iran and the US, as well as the H1N1 outbreak worldwide. Within this framework, in order to detect the Big Bang of an epidemic we require two types of data: A) A snapshot of the active infected cases in each subpopulation during the linear phase. B) A coarse-grained representation of inter-population mobility. Also even with access to only type A data, we can still demonstrate the universal contagion geometric pattern. Additionally, we can estimate errors and assess the precision of the estimations. This comprehensive approach enhances our understanding of when and where epidemics began and how they spread, and equips us with valuable insights for developing effective public health policies and mitigating the impact of infectious diseases on populations worldwide.
\end{abstract}

\section{Introduction} \label{sec:Introduction}
Throughout history, infectious disease outbreaks have significantly impacted human life all over the world \cite{madhav2017pandemics,huremovic2019brief}, causing many deaths \cite{world2008global}. For instance, the COVID-19 pandemic, originating from China \cite{wu2020new}, affected people of all countries \cite{msemburi2023estimates,wang2022estimating}, mentally \cite{pfefferbaum2020mental,galea2020mental,talevi2020mental}, financially \cite{ashraf2020economic}, and beyond.
Human mobility is a key factor in this regard \cite{dalziel2013human,belik2011natural,brockmann2013hidden,balcan2009multiscale,barbosa2018human,anderson1991infectious}, accounting for the spatial spread of diseases, facilitating their propagation through the densely connected networks of global, national, local displacement \cite{antras2020globalization,vespignani2010multiscale,gomez2018critical}. The complex network of travel routes \cite{barabasi1999mean} provides numerous direct and indirect pathways for disease transmission at various scales, with air travel playing a crucial role in the rapid spread of viruses including SARS-CoV-2 \cite{adiga2020evaluating,colizza2006role,lawyer2015measuring,anderson2020will,meidan2021alternating} at the macroscopic level, given their potential to connect distant locations \cite{arino2021describing}. This underscores the importance of taking immediate non-pharmacological interventions \cite{liu2020secondary}, such as air travel restrictions \cite{adiga2020evaluating,arenas2020modeling}, to control disease spread. Understanding the underlying mechanisms of disease transmission at a coarse-grained level, where the mobility network can be considered as a meta-population network \cite{arino2017spatio,bichara2018multi}, is essential for developing effective strategies to control pandemics and save lives. More specifically, answering questions like “Where is (are) the source (sources) of an epidemic?”, “When did it begin?”, which we refer to as the \textbf {Big Bang} of the epidemic, and “How does the outbreak spread through other sub-populations?” after the Big Bang, is of paramount importance.
 
In recent studies, a variety of mathematical models \cite{pastor2015epidemic}, ranging from stochastic models \cite{mollison1977spatial,colizza2008epidemic}, spatio-temporal spreading models \cite{robertson2019spatial,mollison1985spatial}, reaction-diffusion \cite{rass2003spatial}, to agent-based \cite{grenfell2001travelling,merler2010role,balcan2009seasonal,eubank2004modelling,fumanelli2012inferring} and network  \cite{arino2021describing} models and meta-population models \cite{balcan2009seasonal,bajardi2011human,viboud2006synchrony,balcan2012invasion,colizza2007invasion,colizza2008epidemic,arino2015epidemiological,arino2003multi,arino2005multi,bock2019optimal,gaythorpe2016disease,glass2013eliminating,harvim2019transmission,kim2017assessment,lee2015role,matthews2003neighbourhood,arino2020investigation} have been developed to investigate various aspects of disease-spread phenomena. More specifically, \textbf {Effective Distance (ED)} can effectively address the issues stated above \cite{gautreau2007arrival,brockmann2013hidden,lawyer2015measuring,iannelli2017effective,zhong2020country,zhong2020country,aleta2020evaluation}. The effective distance between two sub-populations is defined based on the probability of travel through direct or indirect displacement paths between them. Considering the most probable path between two sub-populations, in 2007 Gautreau \textit{et al} \cite{gautreau2007arrival} and later in 2013 Brockamnn and Helbing \cite{brockmann2013hidden} proposed ansatz for calculating effective distance, and showed a relatively high correlation between ED and the first arrival time of a disease in simulations using the worldwide air transportation network. The methodology was later improved by adding the effects of all possible paths \cite{gautreau2008global}, resulting in a substantial increase in the correlation between the first arrival time and ED. Also this approach has been validated with empirical data of the 2003 SARS and the 2009 H1N1 pandemics\cite{brockmann2013hidden}. Other successful modifications on ED have also been reported \cite{lawyer2015measuring,iannelli2017effective,zhong2020country}. For instance, Zhang \textit{et al} introduced Country Distancing \cite{zhong2020country}, which is similar to the equivalent resistance defined for parallel resistors in electrical circuits. The idea of ED was also successfully tested for the Covid-19 pandemic on the world air traffic data \cite{aleta2020evaluation}. Furthermore, the ED method suggests a technique to identify the source of an epidemics by pinpointing the source that exhibits the highest correlation between arrival time and Effective Distance \cite{brockmann2013hidden}.

 There are also many other methods developed to identify the source of an outbreak in a variety of networks such as a meta-population network or a network of individuals, and in different contexts like disease spreading \cite{tang2018estimating,li2019locating,antulov2015identification,wang2020locating,lokhov2014inferring,prakash2014efficiently,choi2020epidemic}, information spreading \cite{louni2014two,lind2007spreading,altarelli2014bayesian}, food contamination \cite{horn2019locating,horn2017network,schlaich2020gravity}, rumors \cite{ji2017algorithmic,shelke2019source,ji2017algorithmic,seo2012identifying,wang2014rumor,jiang2016rumor,shah2011rumors}, diffusion processes on networks \cite{hu2018locating,hu2018localization,hu2019locating,shen2016locating,paluch2018fast,comin2011identifying}, etc. Typically, the aim of these studies is to identify the source of spread from a “snapshot”, for example number of infected ones, which is the state of the system after the start of the spread.

Despite the success of ED methods, there are some limitations. Some examples follow. First, their definitions often rely on intuition rather than being grounded in comprehensive mathematical models, which may hinder their clarity and interpretability, as well as impede a deep understanding of the contagious dynamics they aim to describe. Second, it is usually assumed that the disease originates from a specific location in the network (the source) and contaminates other nodes over time. However, this assumption is not necessarily true. For instance, on a country scale, the disease can reach different nodes (states or provinces) from outside the network during its spread, effectively acting as multi-sources within the country. Finally, the methods lack any correction of time to detect the beginning of the epidemic in data or any error analysis to check the validity of the spatial and temporal source estimations.

In this paper, we address these gaps by first developing a mathematical framework based on intra-population SIR model dynamics and inter-population mobility within a metapopulation network. We then derive expressions to identify the source or sources of any given epidemic, as well as its starting time, given the number of infected individuals and coarse-grained mobility data at a specific time. Additionally, we propose a novel definition for ED, which \textbf{universally} relates the overtaking time of nodes to their ED from the source of the epidemic. We validate our method using empirical data from the COVID-19 pandemic in Iran and the United States, as well as H1N1 worldwide.

\section{Our Mathematical Framework}
\label{sec:Our General Mathematical Framework} 

\begin{figure}[ht]
\includegraphics[width=0.98\textwidth]{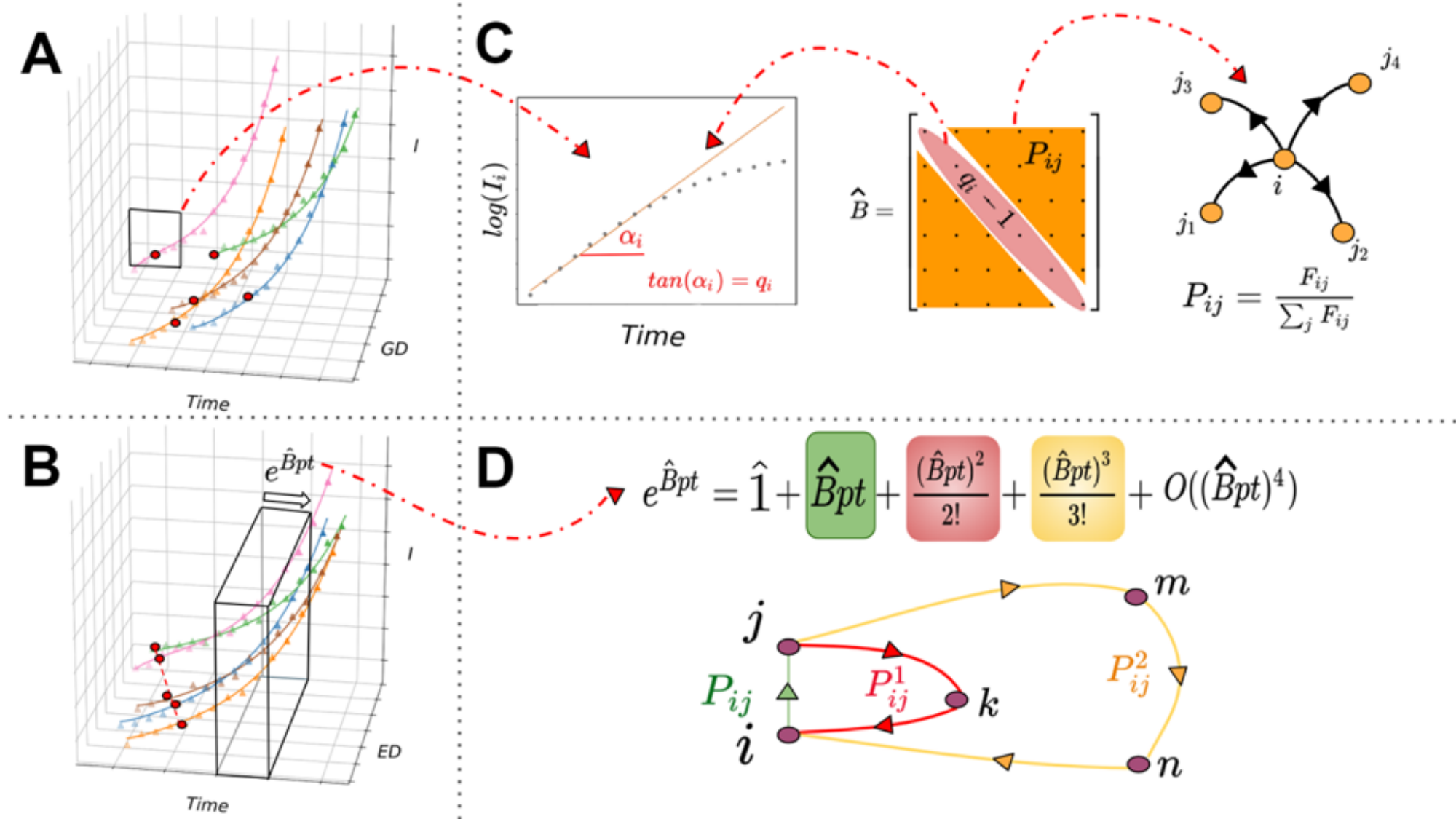}
\caption{\textbf{(A)} The number of infected people in different nodes ($\Vec{I}(t), $\ref{eq: 1-screen shot}) in a network versus time. The third dimension represents the geographical distance from the source in this specific network and it does not show any pattern. \textbf{(B)} The same number of infected people in different nodes in a network versus time; but this time they are plotted based on their effective distances from the source. This time a linear relation between the overtaking times and \textbf{Effective Distances} can be seen. The evolution of the number of infected people is given by the operator, $e^{\hat{B}pt}$, see Eq. \ref{eq: 1-evolution}. \textbf{(C)} \textbf{Center}: The matrix $\hat{B}$ is defined based on the transition probability (see S.M. \ref{SM-sec:Definition of flow matrix and Probability Matrix}) and the slope of the linear part of the SIR dynamics (see S.M. \ref{SM-sec: Driving exponential behavior of I(t) in the early stage of dynamic}). \textbf{left}: $\log I_i$ vs time is illustrated. $q_i$ is the slope of the linear part. The behavior of $\log(I_i)$ in the early stage of the dynamic is linear and the slope of this line is $q_i$. \textbf{Right}: the transition probability is made from the flow between node i and j. \textbf{(D)} When $e^{Bpt}$ is expanded several terms are generated, each containing a power of matrix P. Different terms generate the probability of intermediary transitions. For example, the $\kappa_{th}$ term of the expansion corresponds to a path containing $\kappa-1$ intermediary node.}
\label{fig: Introduction}
\end{figure}

Here we aim to find the Big Bang of a given epidemic dynamics, and in particular, we want to understand how an outbreak spread after the Big Bang. In doing so, the first step is to propose our general mathematical framework which the spreading dynamics adhere to.

When studying the spread of infectious diseases at a given coarse-grained scale, the disease can be considered to spread through a meta population network. In this network, the number of infected and infectious individuals at time t and for subpopulation (node) i, $I_i(t)$, can be put into components of a vector we call $\Vec{I}(t)$:
\begin{equation}
     \Vec{I}(t)=(I_{1}(t), I_{2}(t), ... , I_{n}(t)).
     \label{eq: 1-screen shot}
 \end{equation}

In general, during epidemic dynamics, there is no specific pattern in the number of infected people across all nodes. However, typically, the number initially increases to reach a peak, followed by a subsequent decline, see Fig.~\ref{fig: Introduction}(A). In our following framework, our primary objective is to demonstrate the evolution of $\Vec{I}(t)$ and subsequently discern a straightforward geometric pattern that illustrates the progression of the outbreak within nodes of the metapopulation network. Therefore we show that vector $\Vec{I}(t)$ evolves in time as follows, see Fig.~\ref{fig: Introduction}(B) and the Supplementary Material (S.M.), Section \ref{SM-sec:General Mathematical Framework} for more details: 

\begin{equation}
\Vec{I}(t) = e^{\hat{B}pt} \Vec{I}(0).
\label{eq: 1-evolution}
\end{equation}

The above equation is the solution of the following equation, assuming $S_i \approx N_i$ for the early stages of the dynamic, see S.M. \ref{SM-sec: Driving exponential behavior of I(t) in the early stage of dynamic} for more details:

\begin{equation}
\frac{dI_i}{dt} = [\beta_i S_i I_i -\gamma_i I_i ] +  [p\sum_{j} P_{ji} I_j -p I_i].
\label{eq: 1-dynamic}
\end{equation}
While this equation represents the dynamic of vector $\Vec{I}(t)$ and has two parts:
\begin{enumerate}
    \item spreading within the population, namely \textbf{intra population} (First Bracket).
    \item spreading between nodes, namely \textbf{inter population} (Second Bracket).
\end{enumerate}

For the first part of the equation,  we use a simple Susceptible-Infected-Recovered model (SIR) with a homogeneous mean field approximation (HMFA). Moreover, in the second bracket, we connect all nodes, i.e. each population via a meta-population network. A detailed explanation of how the intra-population term in Equation \ref{eq: 1-dynamic} is derived can be found in S.M., Sections \ref{SM-sec:Definition of flow matrix and Probability Matrix} and \ref{SM-sec:Deriving the intra population term in Eq.2}.\\

In Eq.\ref{eq: 1-evolution}, matrix $\hat{B}$ keeps all of the information regarding the properties of the dynamics (Fig.~\ref{fig: Introduction}(C), center). 
The diagonal components ($q_i$) represent the properties of internal growth of the disease (Intra Population Dynamics). The population of infected people in each node shows exponential behavior at early stages. So, in the plot of $\log(I)$ vs time, there is a linear pattern for each node at the beginning of the outbreak called "linear phase". In this study, we specifically focus on this part of the dynamics. As it can be seen in Fig.\ref{fig: Introduction}(C), left, the slope of this line for node i is $q_i$,  which sits into the $i_{th}$ diagonal component of matrix $\hat{B}$ (Fig. \ref{fig: Introduction}(C), center), see more details in S.M. \ref{SM-sec:Driving q in SIR}. 

Other components of matrix $\hat{B}$  are called $P_{ij}$, which represent the probability of traveling from node i to j. For calculating the value of $P_{ij}$ we use the flow matrix, $F_{ij}$, which keeps the number of people who travel from node i to j in a specific period. (Fig. \ref{fig: Introduction}(C), Right). Further details about probability and flow matrices are available in the S.M. \ref{SM-sec:Definition of flow matrix and Probability Matrix}. 

Now we can expand the solution (Eq. \ref{eq: 1-evolution}) and write it down as:

\begin{equation}
\Vec{I}(t)= (\hat{1} + \hat{B}pt + \frac{(\hat{B}pt)^2}{2} + ...) \Vec{I}(0)
\label{eq:1-Taylor expansion}
\end{equation}

This expansion contains different powers of matrix $\hat{B}$. Since $\hat{B}$ has $P_{ij}$ on its non-diagonal components, different powers of $\hat{B}$ generate different powers of matrix $P$. As matrix $P$ describes the probability of transition directly from node $i$ to $j$, higher powers of $P$ describe the probability of indirect transition from node $i$ to $j$ through a specific number of intermediary nodes in between. This is a new finding we call \textbf{Intermediate Probability}. As can be seen in Fig.\ref{fig: Introduction}(D), the $k_{th}$ component of the expansion holds the intermediate probability of using $\kappa-1$ intermediary nodes (S.M. \ref{SM-sec:Expansion of and appearance of intermediary nodes}).
\newline

\par
The expanded solution can get simpler even further by focusing on the early stage of the dynamic, i.e. the linear phase, and because it takes more time to transit by indirect paths via intermediate nodes. This means that in the early stages of the dynamics, as we go further into the expansion, terms become smaller. Therefore, we can simplify the dynamic by cutting the expansion up to a certain term, keeping only the first terms. This defines the time scales for our model framework. We need to be in a time range in which the assumption of linear evolution for SIR and expansion's cut work together, or $\tau=min\{\frac{1}{\beta - \gamma}, \frac{1}{p}\}$, which constrains the time scale.

As explained, we aim to focus on the very beginning of the spread process, when everything just started and almost no one was infected, and study the expansion of the number of infected individuals in a specific order from the start, like the idea of the \textbf{Big Bang} in cosmology. In the following section, we will introduce several algorithms to solve the challenges and find the starting time and place as well as the hidden spread mechanism of the disease. 

\newpage
\begin{figure} [htbp]
\includegraphics[width=0.95\textwidth]{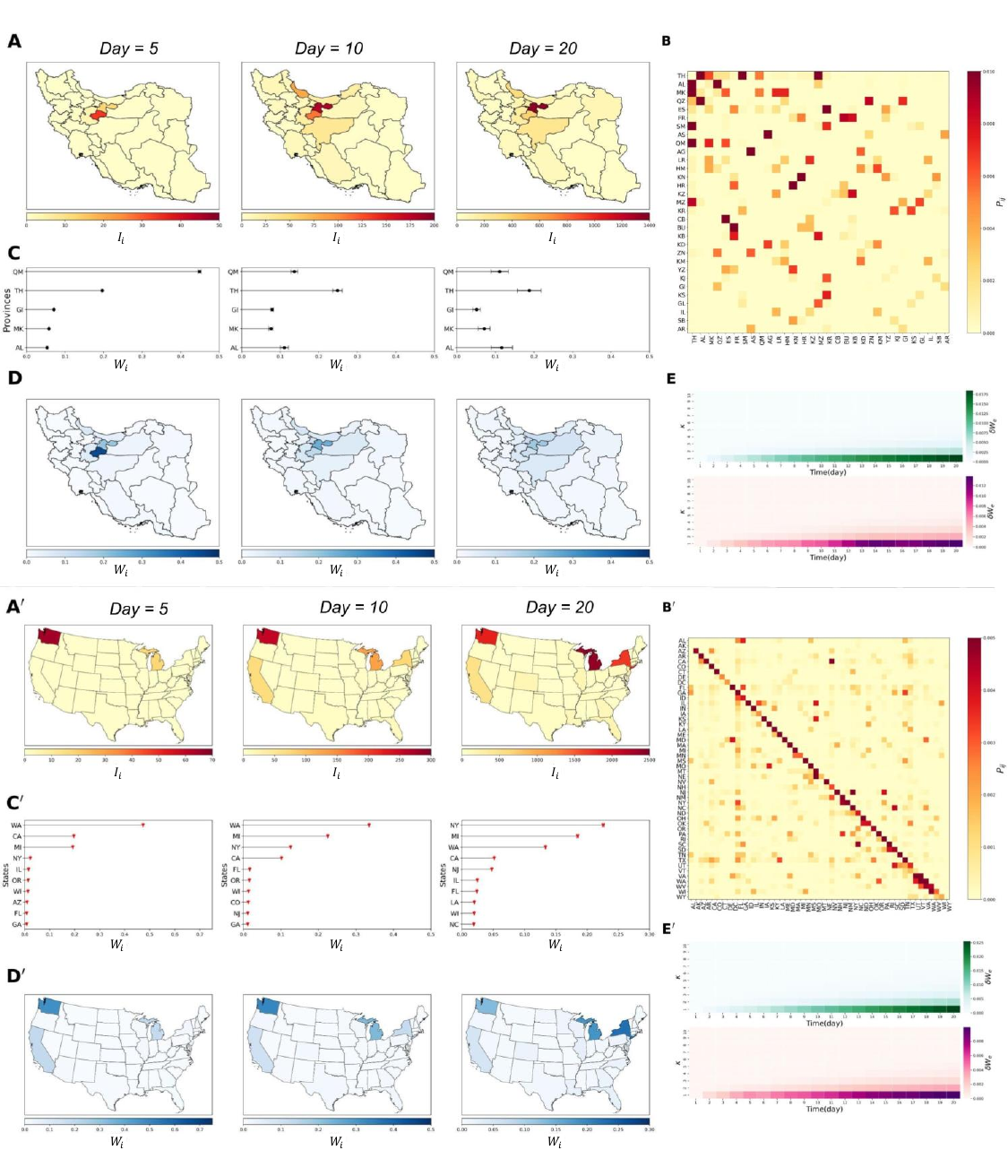}
\caption{\textbf{Where did it start?} 
For better visibility of details, please zoom in.
\textbf{A\&A')} 
In these color maps, the number of COVID-19 infected individuals ($\vec{I^e}(t_e)$) is depicted across various provinces/states of Iran/the United States, captured at different snapshots ($t_e=$ 5, 10, 20 days). The time origin of these snapshots is the official start day of the pandemic in these two countries. 
\textbf{B\&B')} The color map demonstration of the mobility probability matrix, used to calculate the values of $W^e_i$ for Iran/United States. Please refer to the S. M. Section \ref{SM-sec:Data} for more details regarding the data.
\textbf{C \& C')} $W^e_i$ (Eq.\ref{eq: 2-empirical node's weight}), is illustrated for provinces/states of Iran/United States. The black/red points represent the estimated value of $W^e_i$ for each node, and the black/red bars illustrate the value of errors, $\delta W^e_i$. We used the first three terms of the Taylor expansion to calculate weights and errors. Please refer to S.M. \ref{SM-sec:Full details of "Where" Algorithm} for more details.
\textbf{D\&D')} The $W^e_i$ shown in panels C and C' are visualized on geographical maps. 
\textbf{E\&E')} These plots illustrate the reliability of different $\kappa$ values for the cut-off error. The magnitude of $|\Vec{\delta W^e}|$ is depicted in the top color map for the empirical data and in the bottom color map for the simulation results for the networks of Iran and the United States. The vertical axis represents $\kappa$, while the horizontal axis represents time. Here attention should be given to the values of $\kappa=3$ and time (5, 10, 20) used in panels C and C', as they are indicated in both color maps.
We considered $R_0=3$ (Basic Reproduction Number) and $\frac{1}{\gamma}=14(days)$ for all nodes. Please refer to S.M. \ref{SM-sec:Sensitivity Analysis} for sensitivity analysis.}\label{fig: Where}
\end{figure}

\section{Algorithms and Results}
\label{sec:Algorithms and Results} 
\par In this section, using our mathematical framework, we introduce algorithms that reveal \textbf{where} and \textbf{when} the outbreak began and \textbf{how} it spread further using the snapshots of the disease and the flow matrix. In the first algorithm, we detect the potential sources of the disease. In the second one, we estimate the starting time of the spread, then we introduce an algorithm to illustrate a geometric pattern for the spread of the disease. There are different sources of error in the estimations such as approximating the outbreak with SIR model, inaccurate measurement of the number of infected people, and the flow matrix, which is challenging to take into account. Therefore, we only report the theoretical error caused by the cut-off in the expansion of $\exp{\hat{B}pt}$ operator, see details in S.M. Sections \ref{SM-sec:Error in "When" algorithm} and \ref{SM-sec:Error in Effective Distance algorithm}. And the role of estimating epidemiological parameters in error analysis is discussed in the supplementary text (S.M., Section \ref{SM-sec:Sensitivity Analysis}). Given that these algorithms rely on empirical data as input, to distinguish empirical data from theoretical variables, we denote empirical data using the subscript or superscript \textbf{e}. For instance, $\vec{I}(t)$ is a vector containing the number of infected people in our mathematical framework, and $\vec{I}^e(t)$ is the same vector but contains the empirical data of infected people coming from official reports and announcements, see S.M. Sec. \ref{SM-sec:Data}.

\subsection{Where did it start?}
\label{sec: Where did it start ?} 

\par Here, we aim to find the potential sources, \textbf{where} the dynamic began, having $\vec{I}^e(t)$ as empirical data and (Eq.\ref{eq: 1-evolution}) as theoretical formalism. We first develop the theoretical basis of the algorithm and then discuss how to apply it to COVID-19 data of Iran and the USA.

\par
In a network with $n$ nodes, there is a n-dimensional vector space whose i-$th$ basis represents the node i. We define the basis $\vec{i}$ as
\begin{equation}
    \vec{i} = (0,0,...,1,..0),
    \label{eq: 2-basis vector}
\end{equation}
in which its i-$th$ component is 1 and others are zeros.
Now we can expand vector $\vec{I}(t)$ in this space using these bases. The value of the component of this vector on each basis indicates the contribution of that basis to the spread of the disease. As can be seen in Eq. \ref{eq: 1-evolution}, the vector $\vec{I}(t)$ evolves in time and we have to evolve the bases in time as well, to rewind the dynamic to the origin of the time (see S.M. \ref{SM-sec:Full details of "Where" Algorithm}). Therefore, we redefine the bases as follows:

\begin{equation}
    \vec{i'} = e^{\hat{B}pt} \vec{i}.
    \label{eq: 2-basis transformation}
\end{equation}

Now, we define the "weight of a node" as:
 
 \begin{equation}
    W_i = \frac{\Vec{i'}.\Vec{I}(t)}{\sum_{\Vec{i'}}{\Vec{i'}.\Vec{I}(t)}},
    \label{eq: 2-node's weight}
\end{equation}
in which $W_i$ is a number between 0 and 1 that shows the impact and contribution of the node i on the spreading and evolution of the disease at any time within the linear range of the dynamic. If $t$ is measured exactly from the \textbf{origin of time} of the disease, then $W_i$ represents the \textbf{spatial source of the disease}, which can be a single or multiple source. For example, for a given network if $W_i=1$, it means that node i was the only source of the network. For empirical data we use the vector $\vec{I^e}(t_e)$ in Eq. \ref{eq: 2-node's weight}, instead of vector $\vec{I}(t)$. $t_e$ is time, measured from the officially reported temporal origin of the disease. Therefore
 \begin{equation}
    W^e_i = \frac{\vec{i'}.\Vec{I^e}(t_e)}{\sum_{\vec{i'}}{\vec{i'}.\Vec{I^e}(t_e)}}    \label{eq: 2-empirical node's weight}.
\end{equation}
It is important to note that $W^e_i$ calculation is independent of the structure of the network as there is no constraint (such as topology of the network) on the flow matrix used in the calculation.

In figure \ref{fig: Where}, we implement this method for the empirical COVID-19 data of Iran and the US (Panel A and A') and illustrate the results. Also the mobility probability matrix are plotted respectively in panels B and B'. In the case of Iran (Panel C and D), we observe that the values of weights differ by implementing different snapshots. In the first scenario, Qom has the highest weight value, but when considering data from later days, Tehran (the capital province of Iran) surpasses it. Since the largest international airport in Iran (Imam Khomeini) is located between Qom and Tehran provinces, and other important nodes like Gilan and Mazandaran (which also have high values of $W$) are geographically close to this airport, we can conclude that it is more probable that the very first seed of the disease came from this airport to the country and Tehran and Qom are the most probable sources of the disease, which is consistent with official reports.
For the case of the US, (C' and D') Washington state has the highest value of weight in the first plot, but over time, based on the values of W, and error bars, other states like Michigan, California, and New York could also be considered important nodes. So, based on this figure, our model predicts that the states of Washington, Michigan, and New York were the most probable sources of the pandemic in The US.

The reasons for these fluctuations can be the uncertainty in empirical data (error in testing, organization error, and other sources of errors), deviation from the assumption of our model (see discussion section), and errors in calibration ($q_i$ and $p$). 

To calculate the error of the Eq. \ref{eq: 2-node's weight} and Eq. \ref{eq: 2-empirical node's weight} we consider the value of \textbf{cut-off error} since only the first terms of the Taylor expansion of $e^{Bpt}$ is used in the calculations. If $W_i^e$ is calculated using the first $\kappa$ terms of the $e^{Bpt}$ expansion, the error of $W_i^e$ can be calculated using $\kappa$+1 term as

 \begin{equation}
\delta W^e_i = \frac{(\frac{t^{\kappa+1}\hat{B}^{\kappa+1}}{\kappa!}\vec{i'}).\vec{I^e}(t_e)}{\sum_{\vec{i'}}\vec{i'} \cdot \vec{I^e}(t_e)}
    \label{eq: 2-node's weight error}.
\end{equation}
\\
The value of error depends both on $\kappa$ and the value of $t_e$. By increasing $\kappa$ or decreasing $t_e$, we expect to get a smaller value of error. To get a better idea from the value of cut-off error in a whole network, we define the \textbf{error of cut-off vector} as:

 \begin{equation}
    |\Vec{\delta W^e}| = 
    \frac{\sqrt{\sum_i{(\delta W^e_i)^2}}}{n},
    \label{eq: 2-error cut-off vector}
\end{equation}
which is shown versus $\kappa$ and $t_e$ for Iran (panel E) and the US (panel E').

\subsection{When did it start?}
\label{sec:When did it start ?} 
\begin{figure}[ht]
\centering
\includegraphics[width=\textwidth]{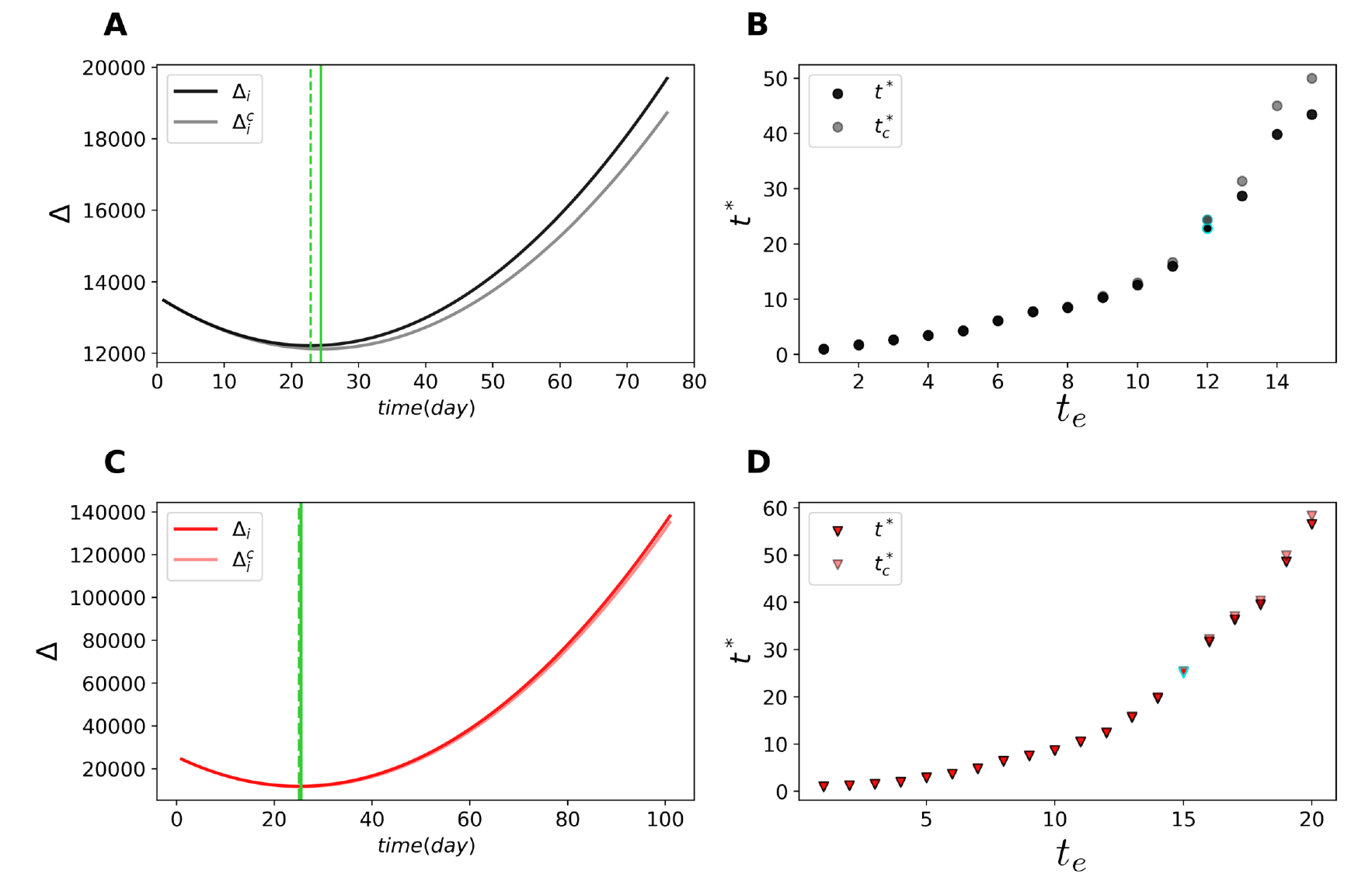}
\caption{\textbf{When did it start?} \textbf{A)} The value of MSE (Eq.\ref{eq: 3-MSE}) is illustrated versus time (the black curve) for Qom province using a snapshot for the \textbf{$12_{th}$} day of the COVID-19 pandemic in \textbf{Iran}. As shown, it has a minimum at ($t_i^*$) (Eq.\ref{eq: 3-starting time}). When the third term of the expansion is added to the calculation, $\Delta_i$ transforms to the gray curve ($\Delta_i^c$), with the minimum shifted approximately two days. \textbf{B)} The estimated origin of time ($t^*_i$) versus the date of the snapshot ($t_e$) for Iran, considering Qom as the origin node. The points highlighted in green show the used snapshot in panel \textbf{A}. \textbf{C)} The value of MSE (Eq.\ref{eq: 3-MSE}) is illustrated versus time (the red curve) for the Washington state using a snapshot for the \textbf{$15_{th}$} day of the COVID-19 pandemic in \textbf{the United States}. It has a minimum at ($t_i^*$) (Eq.\ref{eq: 3-starting time}). When the third term of the expansion is added to the calculation, $\Delta_i$ transforms to the pink curve ($\Delta_i^c$), which approximately lies on the red curve. \textbf{D)} The estimated origin of time ($t^*_i$) versus the date of the snapshot ($t_e$) for the US, considering Washington as the origin node. The points highlighted in green show the used snapshot in panel \textbf{C}. Please refer to the S. M. Section \ref{SM-sec:Data} for more details regarding the data.}
\label{fig: when}
\end{figure}

In this section, our goal is to estimate the temporal origin of the outbreak. As we already mentioned, the real origin of time may differ from the one that is officially reported. To find the temporal origin we compare the estimated ($\vec{I}(t)$) and reported ($\vec{I^e}(t_e)$) number of infected people at the time $t$ and $t_e$, respectively. We find the temporal origin so that it minimizes the mean squared error (MSE) between $\vec{I}(t)$ and $\vec{I^e}(t_e)$,

\begin{equation}
    \Delta_i (t)=\frac{\sum_{j=1}^n (I_j(t) - I_j^e(t_e)^2}{n}.   
    \label{eq: 3-MSE}
\end{equation}
To estimate the number of infected people, $\Vec{I}(t)$, we assume the source is the node $i$ found by the \textbf{Where} algorithm, Sec. \ref{sec: Where did it start ?}. Eq. \ref{eq: 3-MSE} has a unique minimum at time $t^*_i$ (see S.M. \ref{SM-sec:Full details of "When" Algorithm} and (Fig. \ref{fig: when}))
: 
\begin{equation}
     t^*_i= \frac{1}{i_0} \frac{-\eta(i_0 - I_i^e(t_e)) + p \sum_j(P_{ij}I_{j}^e(t_e))}{\eta^2 + p^2 \sum_j P_{ij}^2},
\label{eq: 3-starting time}
\end{equation}
in which  $\eta = (\beta_i N_i - \gamma_i - p)$, and $i_0$ is the initial number of the infected people in the source and at $t^*_i$ (the estimate of the model for the temporal origin of the outbreak). For the error, we consider the cut-off error by adding the third term of the Taylor expansion as the error-making term to our calculation which shows how the predictions degrade (S.M. \ref{SM-sec:Error in "When" algorithm}).

Fig. \ref{fig: when} demonstrates the result of our algorithm applied to the empirical data of Iran and the US. In panel \textbf{A} and \textbf{C}, $\Delta$(Eq.\ref{eq: 3-MSE}) is shown versus time for Iran and the US, respectively. A minimum exists in both cases, as we showed in Eq.\ref{eq: 3-starting time}. By adding the third term of the Taylor expansion (Eq.\ref{eq:1-Taylor expansion}) to the calculation, the corrected MSE (specified with different colors) shifts to a new curve and the minimum moves a bit (around two days for Iran and less than a day for the US). To understand the accuracy of the algorithm, we illustrate  $t^*_i$ versus $t_e$ for Iran (panel \textbf{B}) and the US (panel \textbf{D}) using snapshots from various days. A linear behavior can be observed up to a certain point in both cases, which indicates the linear range of the dynamics. It is worth noting again that $t^*_i$ describes the starting time of the disease from the epidemiological point of view, while $t_e$ refers to the starting time the official reports claim. So, a difference between these two origins of time is expected.
By utilizing these specific snapshots, the onset of the COVID-19 pandemic is estimated to be 8 February 2020 for Iran and 12 February 2020 for the US, marking the commencement of the widespread outbreak in Iran and the US.

\subsection{How does it spread? The universal pattern of any outbreaks}
\label{sec:How did/does it spread ?} 

In previous sections, we estimated the origin of the disease, trying to answer when and where it began. In this section, we aim to illustrate the simple geometric patterns behind the dynamics.

For the first step, let us define the \textbf{overtaking time} in our mathematical formalism. This is the time when enough number of infected passengers arrive in a susceptible node so that we can consider this node as infected. We assume that it happens when the intra-population spreading in this node (the first bracket in Eq. \ref{eq: 1-dynamic}) becomes greater than the inter-population spreading (the second bracket in Eq. \ref{eq: 1-dynamic}), which means

\begin{equation}
    (N_j\beta_j - \gamma_j)I_j = p (\sum P_{kj}I_k - I_j).
    \label{eq: 4-condition}
\end{equation}
Using the above condition and the dynamic of our model given by Eq. \ref{eq:1-Taylor expansion}
, one can show that the overtaking time is (see S.M. Sections \ref{SM-sec:Details of "Effective Distance" Algorithm} and \ref{SM-sec:Error in Effective Distance algorithm} for more details)

\begin{equation}
t_{O}^{j} = \frac{1}{p} \frac{1}{(2+q_i-q_j)-\frac{P_{ij}^1}{P_{ij}}},
\label{eq: 4-overtaking time}
\end{equation}
in which the overtaking time, $t_O^j$, has been calculated for the node j, given the single source, node $i$, in the network.

To detect a simple \textbf{geometric pattern} of the spread dynamic, as shown in Fig. \ref{fig: Introduction}.b, we define an \textbf{Effective Distance} between a non-origin node j and the single origin node i so that there is a \textbf{linear relation} between overtaking time (Eq. \ref{eq: 4-overtaking time}). Therefore, our effective distance is defined as

\begin{equation}
D_{ij} = \frac{1}{(2+q_i-q_j)-\frac{P_{ij}^1}{P_{ij}}},  
\label{eq: 4-linear relation}
\end{equation}
where
\begin{equation}
D_{ij} = pt_O^{j}.  
\label{eq: 4-effective distance}
\end{equation}

Our innovative approach to defining effective distance distinguishes itself from previous methods \cite{gautreau2007arrival,gautreau2008global,brockmann2013hidden}. While maintaining a similar geometric pattern of spread, our method uniquely utilizes overtaking time rather than arrival time. Remarkably, we reveal a \textbf{universal} behavior, characterized by a consistent \textbf{slope of one} when plotting $D_{ij}$ against $pt^j_O$, regardless of network or disease characteristics. In the above equation, $p$ represents the inter-population speed of the disease spread among the nodes in a network. 

Our proposed effective distance becomes simpler for some special cases. For example, if all nodes have the same value of q, the effective distance simplifies as 

\begin{equation}
D_{ij} =  \frac{1}{2-\frac{P_{ij}^1}{P_{ij}}},
\label{eq: 4-simple overtaking time}
\end{equation}
which is independent of the disease properties and only depends on the the travel flow. The defined effective distance only includes those nodes that satisfy $\frac{P_{ij}^1}{P_{ij}}<2$ as effective distance and the overtaking times should be positive for the non-origin nodes.

\begin{figure}[ht]
\includegraphics[width=\textwidth]{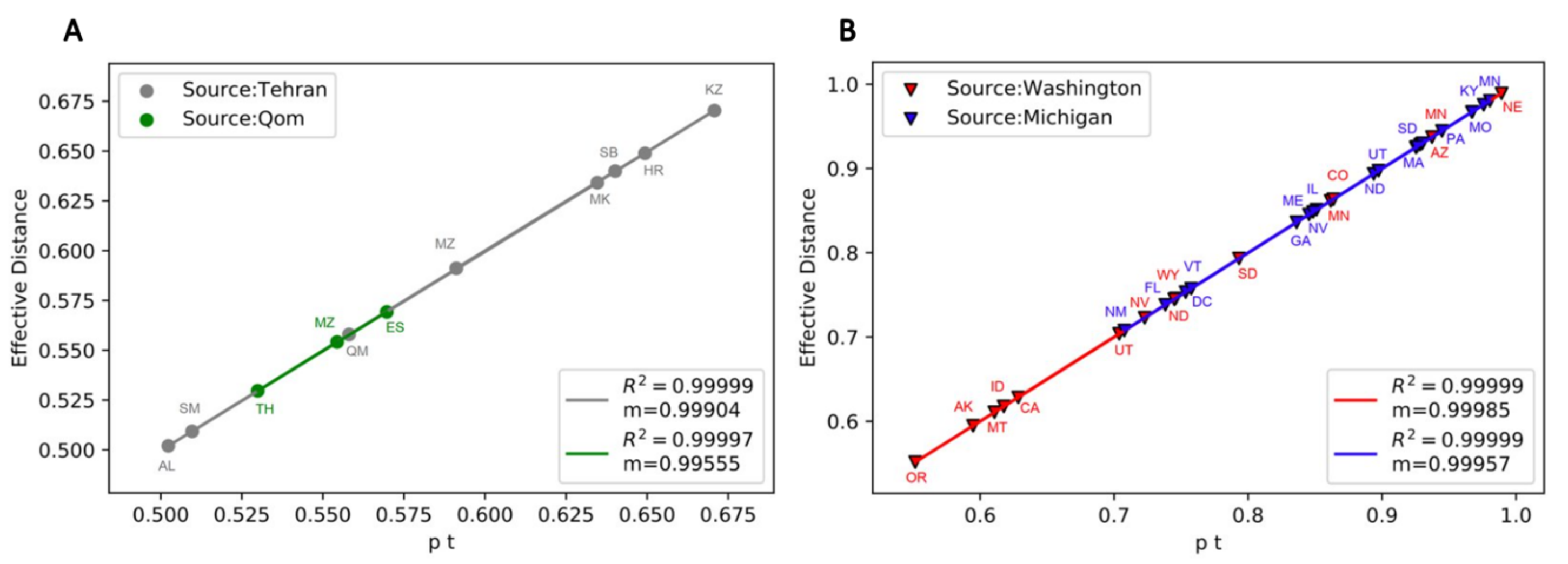}

\caption{\textbf{Effective distance vs overtaking time (simulation result)}: The effective distances are illustrated versus $pt^j_O$ (Eq.\ref{eq: 4-linear relation}) for the empirical mobility data of Iran (Panel A) and the United States (Panel B). Each panel represents two scenarios: we put the initial seed of the disease on a different node in each scenario. These two nodes are selected from the nodes with the higher chance of being the original COVID-19 source in Iran and the US based on the \textbf{Where} algorithm results. In each scenario, we simulated the spread of the disease in the network assuming $R_0=3$ (Basic Reproduction Number) and $\frac{1}{\gamma}=14$ days for all nodes. In panel A, Gray/Green nodes represent the results of the simulation for Tehran/Qom, respectively as the source nodes. Also, the gray and Green lines are the best-fitted lines to the data, with their slope and regression shown in the legend. In Panel B, Red/Blue nodes represent the simulation results for Washington/Michigan as source nodes, with the Red and Blue lines showing the best-fitted lines to the data. Please refer to the S. M. Section \ref{SM-sec:Data} for more details regarding the mobility data.}
\label{fig: Effective distance}
\end{figure}

Fig. \ref{fig: Effective distance} shows the result of the effective distance analysis in two panels. In both panels, the spreading of the disease has been simulated with the SIR model for meta-population networks using the empirical mobility data of Iran (panel A) and the empirical mobility network of the United States (panel B). In each panel, the simulation has been repeated twice (Green and Gray in the left panel, Blue and Red in the right panel), each time with a different source node. The \textbf{Where} algorithm is used in the specific choice of the source nodes. As shown, there is a linear relation between the defined effective distance and $pt$, with the universal slope of one. Changing the source node does not change the linear relation and the value of the slope.

Implementing effective distance analysis with empirical data can pose several challenges. Some of these are outlined below. First, what is reported as the arrival time in official data is not necessarily the same as what we defined as overtaking time in Eq. \ref{eq: 4-overtaking time}, even though they are close. Second, measuring the exact value of the mobility probability matrix is difficult, especially due to the intervention policy in each region at the beginning. Finally, the initial number of infected people ($i_0$) is not necessarily known.
 
It is possible to overcome the challenges stated above by estimating the effective distance of the node $j$ from the source using the number of infected people in that node ($I^e_j(t_e)$). Using the mathematical framework, one can show that $I^e_j(t_e)$ can be estimated by a parabola (see S.M. Eq. \ref{SM-eq:parabola}) in the short enough period of time after the arrival of the disease to the node. Assuming the overtaking time occurs no longer after the arrival time, we estimate the effective distance of the node $j$ from the source only by the found parameters from the fit (S.M. \ref{SM-sec:How to estimate Effective Distance without mobility data}) 

\begin{figure*}[ht]
\centering
\includegraphics[width=\textwidth]{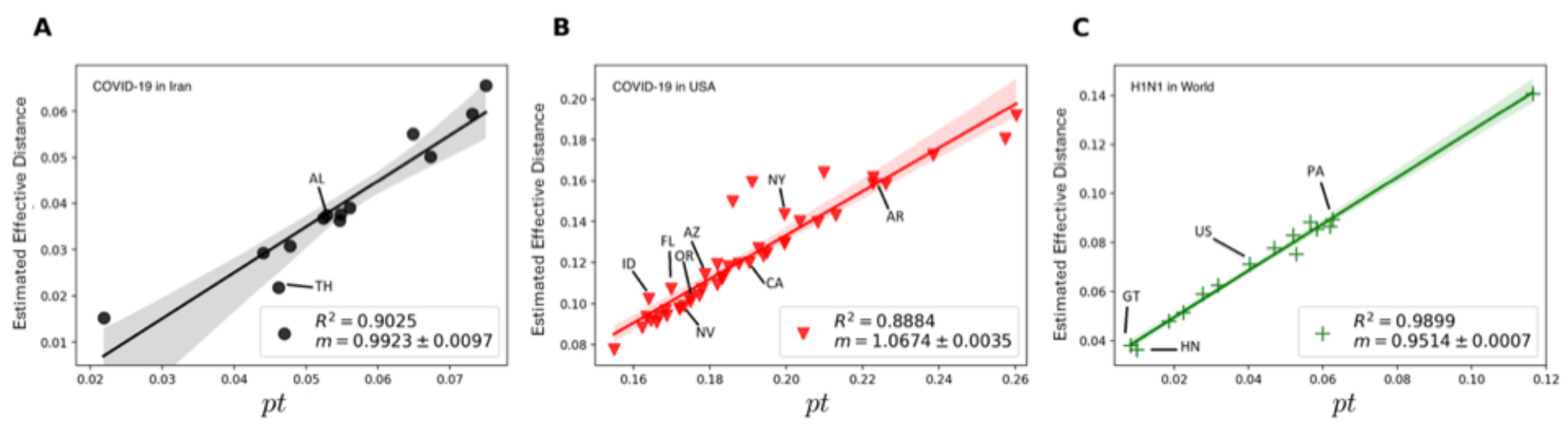}
\caption{\textbf{Estimated effective distance vs overtaking time (empirical data result)}: The estimated effective distance (S.M. \ref{SM-sec:How to estimate Effective Distance without mobility data}) versus the scaled overtaking time ($pt$) is illustrated for \textbf{A}) the Covid-19 pandemic in Iran, \textbf{B}) the Covid-19 pandemic in the United States, and \textbf{C}) H1N1 pandemic in the world (2009), based on the empirical data ($\vec{I^e}(t_e)$) of the pandemics. Please refer to the S. M. Section \ref{SM-sec:Data} for more details regarding the data.}
\label{Fig: effective distance-empirical}
\end{figure*}

Fig. \ref{Fig: effective distance-empirical} shows the estimated effective distance and overtaking time for the empirical data of the COVID-19 pandemic in Iran (panel A), the US (panel B), and the H1N1 pandemic in 2009 in the meta-population of the world. As shown,  there is a linear relation between effective distance and overtaking time in all instances, with the universal slope close to 1. This result demonstrates that the linear relation with a slope of one remains independent of both the disease type and the size and structure of the meta-population network in which the disease spreads.    

\section{Concluding Remarks}
\label{sec:Concluding Remarks} 

In summary, we introduced a mathematical framework based on the SIR model for meta-population networks, incorporating inter-population mobility. We derived a compact equation (Eq.~\ref{eq: 1-evolution}) that represents the time evolution of the number of infected individuals using the mathematical operator $e^{\hat{B}pt}$. We showed how different terms in the Taylor expansion of the operator represent possible transmission paths with different number of intermediary nodes. Based on this general mathematical framework and the provided data, we were able to determine where and when the outbreak began, as well as how it spread within the meta-network.

Firstly, we derived a measure indicating the contribution of each node to disease spread, whether in single-source or multi-source pandemics. Our analysis of COVID-19 revealed that Qom, Tehran, Gilan, and Mazandaran carry the greatest weight in Iran, indicating these provinces as probable sources of the pandemic. This observation aligns with the proximity of these provinces to Imam Khomeini International Airport and the relatively high volume of travel to these areas. Likewise, Washington, Michigan, New York, and California were identified as likely sources of the pandemic in the US.   

Secondly, we derived an expression to find the temporal origin of a pandemic. Thus, we estimated the beginning date of the COVID-19 pandemic in Iran and the US is Feb. 8, 2020, and Feb. 12, 2020, respectively. These dates precede the officially announced start dates in both countries, suggesting that the pandemic may have begun earlier than previously thought. 

Thirdly, we introduced a novel definition for Effective Distance and demonstrated that the effective distance of a node from the source exhibits a linear relationship with the scaled overtaking time ($pt$), characterized by a universal slope of one. Importantly, this relationship remains independent of the epidemiological parameters of the disease and characteristics of the meta-population network, such as the number of passengers and network structure. This assertion is supported by our simulation results for Iran and the US. Finally, we showed how the effective distance can be estimated only with the data of the number of infected ones in the network. We applied this method to the data from the COVID-19 pandemic in Iran and the US, as well as the 2009 H1N1 pandemic. Our analysis confirmed the existence of a linear relationship with the universal slope of one.

Combining all reported observations, our analysis underscores the following practical implications: Given that the speed of disease propagation in the network is directly proportional to travel probability, $p$, this emphasizes the crucial role of implementing travel restrictions during the early stages of a pandemic. Additionally, our findings highlight the importance of predicting more accurately when and how diseases reach the next node. This insight provides policymakers with a better understanding of the optimal strategies for implementing lockdowns or travel restrictions, thereby effectively mitigating the spread of infectious diseases. 

Our work presents several theoretical implications and prospects for the research community. Firstly, unlike similar studies \cite{gautreau2007arrival,gautreau2008global,brockmann2013hidden}, our definition of effective distance in this paper is directly derived from the mathematical model that describes the phenomenon, rather than relying solely on intuition or data analysis. Additionally, our analysis reveals that effective distance exhibits a universal geometric pattern, contributing to a deeper understanding of epidemic dynamics across different contexts. Secondly, our study addresses fundamental questions such as where and when pandemics begin within a coherent mathematical framework, shedding light on essential aspects of disease spread. 
However, our method has limitations stemming from the simplifying assumptions we made. Firstly, we utilized the SIR model for meta-population networks, which can be extended by incorporating more complex epidemiological models, see S.M. Section \ref{SM-sec:Using SEIR instead of SIR} as an example. Secondly, we treated certain parameters as fixed, which may not always hold true. For instance, we assumed that the number of susceptible individuals remains constant and equal to the node’s population at the early stage of the dynamic. Additionally, we supposed that $\gamma$ is constant across nodes and that the flow matrix remains fixed over time. While these assumptions are reasonable in many cases, they may not accurately reflect reality in all scenarios. Furthermore, as demonstrated, systematic errors can arise from ignoring higher-order terms of the Taylor expansion (Eq.~\ref{eq:1-Taylor expansion}). Therefore, our algorithms and results can be enhanced by avoiding mathematical simplifications and improving data quality. Each of these aspects warrants further investigation in future studies.

\section{Acknowledgement} The authors would like to acknowledge Yamir Moreno for his valuable comments and express gratitude to Hossein Afshin for providing us with Iran's passenger flow data.
\color{black}

\newpage
\bibliographystyle{plain}
\bibliography{References.bib}

\newpage
\section{General Mathematical Framework}
\label{SM-sec:General Mathematical Framework}

\subsection{Notation of variables and parameters}

\label{SM-sec:NotationParameters}

\begin{table}[ht]
\begin{tabular}{ |p{1.6cm}||p{3cm}|p{1cm}|p{9cm}|  }
 \hline

 Parameter  & Name &Unit &Definition\\
 \hline
$\beta$   & Transmission rate    & $\frac{1}{day}$&  Transmission rate in the SIR dynamics\\
$\gamma $& Recovery rate    & $\frac{1}{day}$&  Recovery rate in the SIR dynamics\\
$R_0 $& Basic reproduction rate    & -&  The transition rate over the recovery rate\\
$q_i$   & -     & -&   the slope of the linear part of the dynamics for node $i$, see sections \ref{SM-sec: Driving exponential behavior of I(t) in the early stage of dynamic} and \ref{SM-sec:Driving q in SIR}. \\
 \hline
$S_i $& Susceptible in the sub-population $i$    & -&  Number of susceptible individuals in node $i$ \\
$I_i $&   
Infected people in node $i$   & -&   Number of infected/infectious individuals in node $i$ \\
$I_i^e $&  Reported infected individuals in node $i$  & -&   Number of infected/infectious individuals in node $i$ reported by  (empirical data) \\
$\Delta_i $ & MSE  & -&  Mean squared error between $I_i$ and $I_i^e$\\
$i_0 $&  Total initial patients  & -&  The total number of patients at the beginning of the dynamic (sum over $\vec{I}(0)$). \\
$R_i $&   Recovered people in a sub-population $i$  & -&   Number of recovered individuals in node $i$ \\
 \hline
 $N$&  Total population  & -& The total population studied within the meta-population \\
$N_i$&  Population of the sub-population $i$ & -& - \\
$n $& Number of sub-populations  & -&  -\\

$N_p$&  Total passengers population  & -& Number of daily travelers\\
 \hline
$F_{ij}$&  Flow per unit time & $\frac{1}{day}$&Number of daily travelers from the node $i$ to the node $j$\\

$P_{ij}$& Passenger probability per unit time   & -& Probability that a traveler travels from the node $i$ to the node $j$\\
$p$ &$ \frac{N_p}{N}$ &  $\frac{1}{day}$& Travel probability of an individual between sub-populations \\
 \hline

$\hat{B}$ &  Time evolving operator  & -& See the main text\\

$D_{ij}$&Effective distance between node $i$ and $j$ & -& see Eq. \ref{eq: 4-effective distance}\\
$t_o^{j}$&  Overtaking time of node j  & - & When the intra-population dynamics surpass the inter-population dynamics at node j, see Eq. \ref{eq: 4-overtaking time} \\
 \hline

\end{tabular}
\caption{All used parameters in our mathematical framework are brought here.}
\label{SM-table: used parameters}
\end{table}
\newpage

\subsection{Definition of flow matrix and mobility probability Matrix}
\label{SM-sec:Definition of flow matrix and Probability Matrix}

$P_{ij}$, is defined as the probability that a person travels from node $i$ to node $j$. Also, $p$ is defined as the average travel probability in the network. The \textit{Flow Matrix} is a matrix representing the movements within the network. $F_{ij}$ is the number of travellers from node $i$ to $j$ per time. So, $\sum_{j} {F_{ij}}$ is the total number of passengers exiting from the node $i$. Based on this definition, we make the probability matrix and the parameter $p$ as the following:

\begin{equation}
 P_{ij} = \frac{F_{ij}}{\sum_{j} {F_{ij}}}
\end{equation}

\begin{equation}
 p = \frac{\sum_{j}\sum_{i} {F_{ij}}}{N},
\end{equation} where $N$ is the total population of the network.

\subsection{Deriving the intra population term in Eq. \ref{eq: 1-dynamic}}
\label{SM-sec:Deriving the intra population term in Eq.2}

To derive the number of infected people in node $i$, we have to calculate the flow of infected people in and out of this node. To calculate the number of infected people per time, who leave this node and go to other nodes, first, we can use the concept of flow and calculate the number of infected people that go to node $j$ :
\begin{equation}
F_{ij} \frac{I_i}{N_i}
\end{equation}
Then, we can sum over the index $j$ to calculate the total number of people who leave the node $i$:
\begin{equation}
\sum_{j}F_{ij} \frac{I_i}{N_i}.
\end{equation}
If we consider the travel probability in node $i$ to be

\begin{equation}
p_i = \frac{\sum_{j}F_{ij}}{N_i},
\end{equation}
then we can rewrite the above equation and derive the following differential equation for the outgoing population in node $i$:
\begin{equation}
\frac{dI_i}{dt} = -p_i I_i
\end{equation}

Now to calculate the number of infected people that enter node $i$, we calculate the number of infected people who travel from node $j$ to $i$ per time :

\begin{equation}
F_{ji} \frac{I_j}{N_j}
\end{equation}
Then, if we multiply both the numerator and denominator by 
\begin{equation}
\sum_{i}F_{ji}
\end{equation}
the resulting equation is
\begin{equation}
\frac{F_{ji}}{\sum_{i}F_{ji}}\frac{\sum_{i}F_{ji}}{N_j} I_j.
\end{equation}
The first term on the left side is the definition of a probability matrix,
$P_{ji}$. So the number of infected people who travel from node $j$ to $i$ is
\begin{equation}
    p_j P_{ji} I_j.
\end{equation}
By summing up the above equation over index $j$, we have
\begin{equation}
\sum_{j}p_j P_{ji} I_j.
\end{equation}
Now, the full evolution of infected people in node $i$ is
\begin{equation}
\frac{dI_i}{dt} = -p_i I_i + \sum_{j} p_j P_{ji} I_j.
\label{SM-eq:diffusion0}
\end{equation}

To derive the final formula, we assume the travel probability is the same among the nodes of the network and is equal to the average value of being a traveler in the whole network. Therefore
\begin{equation}
    p_i \approx p = \frac{\sum_{j} \sum_{i} {F_{ij}}}{N},
\end{equation}
which means
\begin{equation}
\frac{dI_i}{dt} = -p I_i + p \sum_{j} P_{ji} I_j.
\label{SM-eq:diffusion}
\end{equation}

\subsection{Deriving exponential behavior of \texorpdfstring{$\Vec{I}(t)$} in the early stage of the dynamic (Eq. \ref{eq: 1-evolution})}
\label{SM-sec: Driving exponential behavior of I(t) in the early stage of dynamic}

Assuming the spread of the disease is at its early stages, we can estimate $S_i$ with $N_i$ and rewrite the evolution equation as
\newline
\begin{gather}
S_i\approx N_i\\
\frac{dI_i}{dt} =(\beta_i N_i  -\gamma_i -p)I_i + \sum_{j} p P_{ji} I_j, 
\label{SM-eq: Model}
\end{gather}
in which $P^{T}_{ij}$ and $q_i$ as, 
\begin{gather}
P^{T}_{ij} = P_{ji}\\
q_i=\frac{\beta_i N_i - \gamma_i}{p}. 
\end{gather}
The equation can be rewritten as 
\begin{equation}
\frac{dI_i}{dt} =p \sum_{j} (P^{T}_{ij} + \delta_{ij}(q_j-1))I_j, 
\end{equation}
where $\delta_{ij}$ is the Kronecker delta and is equal to 1 when $i=j$ and zero when $i!=j$. Considering  $ B_{ij} = P^{T}_{ij} + \delta_{ij}(q-1)$, the above equation can be simplified as
\begin{equation}
\frac{d\Vec{I}}{dt} =p \hat{B} \Vec{I}, 
\end{equation}
in which matrix $\hat{B}$ is

\begin{equation}
\hat{B}=
\begin{bmatrix}
q_1 - 1 & P_{21} & P_{31} & \ldots \\
P_{12} & q_2 - 1 & P_{32} & \ldots \\
\ldots & \ldots & \ldots & \ldots \\
\end{bmatrix}
\end{equation}

The solution to the above differential vector equation is 
\begin{equation}
\Vec{I}(t) = e^{\hat{B}pt} \Vec{I}(0),
\label{SM-eq:tylor}
\end{equation}
 where, $\Vec{I}(0)$ represents the initial vector of patients.

\subsection{Deriving \texorpdfstring{$q_i$} in the early stage of dynamics}
\label{SM-sec:Driving q in SIR}
\par
Assuming the spread of the disease is at early stages in the SIR model, we can estimate $S_i$ with $N_i$ and rewrite the evolution equation as
\newline
\begin{gather}
S_i\approx N_i\\
\frac{dI_i}{dt} =\beta_i N_i I_i -\gamma_i I_i,
\end{gather}
with the solution of 

\begin{equation}
I_i(t) = I_i(0) e^{\frac{\beta_i N_i  -\gamma_i}{p}(pt)}.
\end{equation}
The above equation can be rewritten in the simpler form of

\begin{equation}
I_i(t) = I_i(0) e^{q_i(pt)}.
\end{equation}
As a result, we can expect a linear behavior if we plot $\log I_i(t) $ vs $pt$
\begin{equation}
\log I_i(t) = \log I_i(0) + q_i(pt).
\end{equation}
The slope of this line is $q_i$.

\subsection{\texorpdfstring{$e^{\hat{B}pt}$} expansion and intermediary nodes}
\label{SM-sec:Expansion of and appearance of intermediary nodes}

 Eq. \ref{SM-eq:tylor} can be expanded as
\begin{equation}
\Vec{I}(t) = (\hat{1} + \hat{B}pt + \frac{(\hat{B}pt)^2}{2!} + ...) \Vec{I}(0),
\end{equation}
where $\hat{1}$ is the identity matrix. If we label the source node with $i$
, and the non-source nodes with $j$, for node $j$ the above equation leads to :

\begin{equation}
    P_{ij}i_0 pt + \left[ \frac{i_0 p^2 t^2}{2} \left( (P_{ij})(q_i+q_j-2) + P_{i1}P_{1j} + P_{i2}P_{2j} + \ldots + P_{in}P_{nj} \right) + \ldots \right]
\end{equation}

\par
In the above equation, the term :

\begin{center}
$ P_{i1}P_{1j} + P_{i2}P_{2j} + .... P_{in} P_{nj} =\sum_k P_{ik}P_{kj}$
\end{center}
is the probability that one travels from node $i$ to $j$ via one of the intermediary nodes between them. We name it $P_{ij}^{1}$, therefore:

\begin{equation}
    I_j(t) = P_{ij}i_0 pt + \frac{i_0 p^2t^2}{2} \left( P_{ij}(q_i+q_j-2) + P_{ij}^{1} \right) + \ldots
    \label{SM-eq:non-source}
\end{equation}

For the source node, the evolution equation is
\begin{equation}
    I_i(t) = i_0(1+pt(q_i-1)) + ...
    \label{SM-eq:source}
\end{equation}

\section{Algorithms and Results}
\label{SM-sec:Algorithms and Results}

\subsection{Details of Where algorithm}
\label{SM-sec:Full details of "Where" Algorithm}
Assume one has a snapshot of the disease state reported exactly $t$ days after the beginning of the pandemic:
\begin{equation}
(I_{1}^e, I_{2}^e, ... , I_{n}^e). 
\end{equation}
The goal is to find the sources given the starting time and a snapshot of the disease.\\
We know that the dynamic of $\Vec{I}(t)$  is:

\begin{equation}
    \Vec{I}(t) = e^{\hat{B}pt} \Vec{I}(0)
\end{equation}
If there are some nodes responsible for the spread of the disease in the network with the initial number of patient of $i_{0_i}$, $i_{0_j}$, $i_{0_k}$, \ldots, then $\Vec{I}(0)$ can be decomposed into a summation of different initial vectors, each representing a specific node. Therefore we have 
\begin{equation}
    \Vec{I}(t) = e^{\hat{B}pt} (\Vec{I_i}(0) + \Vec{I_j}(0) + \Vec{I_k}(0) + ... )
\end{equation}
\begin{equation}
    \Vec{I}(t) = \Vec{I_i}(t) + \Vec{I_j}(t) + ... ,
\end{equation}
where
\begin{equation}
    \Vec{I_i}(t) = e^{\hat{B}pt} \Vec{I_i}(0).
\end{equation}

Now if we define the basis vector $\Vec{i}$ to be a vector with 1 in component $i$ and 0 in all other components, we can write:

\begin{equation}
    \Vec{I_i}(0) = i_{0_i} \Vec{i}
\end{equation}

\begin{equation}
    \Vec{I}(t) = (e^{\hat{B}pt} i_{0_i} \Vec{i} + e^{\hat{B}pt} i_{0_j} \Vec{j} + ... ).
\end{equation}
So if we change our basis from $\Vec{i}$ into $\Vec{i'}$ :
\begin{equation}
    \Vec{i'} = e^{\hat{B}pt} \Vec{i},
\end{equation}
it can be easily shown that in the linear range, these new bases are complete and orthogonal.
Now, we define the "weight of a node" as
 
 \begin{equation}
    W_i = \frac{\Vec{i'}.\Vec{I}(t)}{\sum_{\Vec{i'}}{\Vec{i'}.\Vec{I}(t)}}.
\end{equation}
$W_i$ is a number between 0 and 1 and shows the contribution of the node $i$ at the beginning of the spread. If $W_i$ is 1 it means that node $i$ was responsible alone.
So by calculating the defined weight for different nodes, we can understand the role and impact of each node on the spread.

\subsection{Details of When algorithm}
\label{SM-sec:Full details of "When" Algorithm}
Assuming to have a snapshot of the state of the disease at a specific time written down as a vector:
\begin{equation}
    \vec{I^e}=(I_{1}^e, I_{2}^e, ... , I_{n}^e)
\end{equation}
where $I_i^e$ represents the number of infected people in its corresponding node. $n$ is the number of nodes in the network.
We aim to determine the starting time of the disease using both the model and the snapshot.
We define MSE (Mean Squared Error) as a closeness parameter  for two vectors:
\begin{equation}
    \Delta_i (t)=\frac{\sum_{j=1}^n (I_j - I_j^e)^2}{n}         
\end{equation}
in which $i$ is the index of the source and $I_j$ is the number of patients in the node $j$ predicted by the model.
\par
Now, we calculate the closeness parameter of the snapshot vector and the disease vector that comes out of the theory (using the first and the second term) in a determined time $t$:

\begin{equation}
    \Delta_i^{\kappa=1} (t) =\frac{1}{n}((i_0 (1+(\beta N_i - \gamma - p)t) - I_i^e)^2+ \sum_{i=j,j!=i}^{n} (P_{ij}pt i_0 - I_{j}^e)^2 )
\end{equation}

To find the minimum of the closeness parameter, we calculate the derivative of the above equation with respect to the t
($\frac{d\Delta_m(t)}{dt}=0$)

\begin{gather}
 t^*_i= \frac{1}{i_0} \frac{-\eta(i_0 - I_i^e) + p \sum(P_{ij}I_{j}^e)}{\eta^2 + p^2 \sum P_{ij}^2}
\end{gather}
in which  $\eta = (\beta N_m - \gamma - p)$.
\\

It means that the theory predicts the snapshot belongs to t days after the start of the disease. 
So the overtaking time would be t days before the date of the snapshot.

\subsection{Error in When algorithm}
\label{SM-sec:Error in "When" algorithm}
In this subsection we want to explain the way we estimated the error in When algorithm. It is crucial to note that the source of error in this algorithm is coming from the additional term in the Taylor expansion.
In this case, for the source node, the dynamic is:
\begin{equation}
    I_i(t) = C + A_i t + B_i t^2
\end{equation}
And for non-source nodes, the dynamic is :
\begin{equation}
    I_j(t) = A_j t + B_j t^2
\end{equation}
The A,B,C coefficients are constants that are calculated via Eq.\ref{SM-eq:non-source} and Eq.\ref{SM-eq:source}.
Now we define MSE parameter:
\begin{equation}
    \Delta = (C + A_i t + B_i t^2 - I_i^e )^2 + \sum_{j} (A_j t + B_j t^2 - I_j^e)^2
\end{equation}
which leads to the following equation:
\begin{equation}
    \Delta = \Delta_0 + 2t^2 ((C-I_i^e)B_i - \sum_{j} ( I_j^e B_j)t^2)
\end{equation}
By rewriting the right term in the above equation:
\begin{equation}
    \eta = 2(C- I_i^e)B_i - \sum_{j} (I_j^e B_j)
\end{equation}
We are able to rewrite the MSE parameter :
\begin{equation}
    \Delta = \Delta_0 + \eta t^2
\end{equation}
Now, by applying the derivative condition to the new equation :
\begin{equation}
    \frac{d \Delta}{dt} = \frac{d \Delta_0}{dt} + 2 \eta t = 0
\end{equation}
The condition of having zero value derivative happens in:
\begin{equation}
    \frac{d \Delta_0}{dt} = - 2\eta t
\end{equation}
To find the specific point that the above condition is met, we use the Taylor expansion of $\Delta_0$ function in the minimum point of the function($t^*$).
\begin{equation}
    \Delta_0(t) = \Delta_0(t^*) + \frac{1}{2} \frac{d^2 \Delta_0(t)}{dt^2}(\delta t)^2 + ...
\end{equation}
In the above equation, the first derivative term is equal to zero since the $\Delta_0$ function is in its minimum in $t^*$.
Then by substituting the Taylor expansion, we have :
\begin{equation}
    \frac{\delta t}{t} = \frac{\eta}{\frac{d^2 \Delta_0}{dt^2}}
\end{equation}

\subsection{Details of Effective Distance algorithm}
\label{SM-sec:Details of "Effective Distance" Algorithm}

In this part, we define the overtaking time of a disease in a specific node as the time when intra-population dynamics surpass the inter-population dynamic. The critical mathematical condition for this state for node $j$ is :
\begin{equation}
    (N_j\beta_j - \gamma_j)I_j = p (\sum_{k} P_{kj}I_k - I_j)
\end{equation}

If we rewrite this equation, we have:
\begin{equation}
     p(\sum_{k} P_{kj}I_k) - (\beta_j -\gamma_j + p)I_j = 0
\end{equation}
So if we define vector $\Vec{A_i}$ for the j-th node:
\begin{equation}
    \Vec{A_j} = (P_{1j}, P_{2j}, ..., -(q_j+1),...,P_{nj}) 
\end{equation}
In which $-(q_j+1)$ is in the $j_{th}$ component.\\
Now, we can write the critical condition for node $j$ in a simpler way:
\begin{gather}
\Vec{A_j}.\Vec{I} = 0
\end{gather}
If we want to calculate the overtaking time for node $j$, we can use the evolution equation of $\Vec{I}$ :

\begin{equation}
\Vec{I(t)} = \Vec{I(0)} + pt \hat{B}\Vec{I_0}
\end{equation}
If we dot product the vector $\Vec{A_j}$ in both sides of the above equation and set the left side of the equation to be zero :

\begin{equation}
   0 = \Vec{A_j}.\Vec{I_0} + p t_A \Vec{A_j}.\hat{B}\Vec{I_0} 
\end{equation}
So the overtaking time would be:

\begin{center}
$t_{O}^{j} = - \frac{1}{p} \frac{\Vec{A_j}.\Vec{I_0}}{\Vec{A_j}.\hat{B}\Vec{I_0}}$
\end{center}
If we assume the $i_{th}$ Node to be the source, the $\Vec{I_0}$ would be :
\begin{center}
$\Vec{I_0} = i_0(0,0,...1,...0)$
\end{center}
In which the 1 is in the $i_{th}$ component.
Therefore, the numerator of the overtaking time equation would be:  $P_{ij}I_0$ \\
For calculating the denominator, first we have to calculate $\hat{B}\Vec{I_0}$ that would be :

\begin{equation}
    \hat{B}\Vec{I_0} = i_0 (P_{i1},P_{i2},.. ,q_i-1, ..., P_{in})
\end{equation}

In which the $q_i-1$ term is in the $i_{th}$ component.
Now we have to calculate the $\Vec{A_j}.\hat{B}\Vec{I_0}$ that would be:

\begin{equation}
i_0 (P_{i_j}^1 - (2 + q_j - q_i ) P_{i_j})
\end{equation}

In which $P_{ij}^1 = \sum_k P_{ik}P_{kj}$

So by substituting these terms in the overtaking time equation, we have :
\begin{equation}
t_{O}^{j} = \frac{1}{p} \frac{1}{(2+q_j-q_i)-\frac{P_{ij}^1}{P_{ij}}}
\label{SM-eq:overtaking time12}
\end{equation}

\subsection{Error in Effective Distance algorithm}
\label{SM-sec:Error in Effective Distance algorithm}

In this subsection, we aim to explain the way we estimated error in Effective Distance algorithm. The source of error here, is the additional term in the Taylor expansion. So if we keep the second term of the expansion, we have :
\begin{equation}
    \Vec{I(t)} = \Vec{I_0(t)} + pt\hat{B}\Vec{I_0} + \frac{1}{2} (pt)^2 \hat{B}^2 \Vec{I_0}
\end{equation}

Now by applying the condition of overtaking in node $j$:
\begin{equation}
    0 = \Vec{A_j}.\Vec{I_0} + pt_j (\Vec{A_j}\hat{B}\Vec{I_0}) + \frac{1}{2} (pt_j)^2 (\Vec{A_j}.\hat{B}^2 \Vec{I_0})
\end{equation}

By rewriting time we have:
\begin{equation}
    t_j = t^*_j + \delta t_j
\end{equation}

In which, $t^*_j$ is the overtaking time when we only consider the first term of the expansion. hence the equation will be in this form:
\begin{equation}
    0 = \Vec{A_j}.\Vec{I_0} + pt^*_j \Vec{A_j}\hat{B}\Vec{I_0} + p \delta t_j \Vec{A_j}\hat{B}\Vec{I_0} + \frac{1}{2} (p(t^*_j + \delta t)^2 \Vec{A}\hat{B^2}\Vec{I_0}
\end{equation}
The first two terms in the right-hand side of the equation will cancel each other out. By simplification of the above equation the final equation for the error will be :

\begin{equation}
    \frac{\delta t_j}{t^*_j} = \frac{-pt^*_j \Vec{A_j}.\hat{B^2}\Vec{I_0}}{2(pt^*_j \Vec{A_j}.\hat{B^2}\Vec{I_0} + \Vec{A_j}.\hat{B}\Vec{I_0})}
\end{equation}

\subsection{How to estimate Effective Distance without mobility data}
\label{SM-sec:How to estimate Effective Distance without mobility data}

In the previous subsection we introduced a new definition of effective distance and showed its linear relation with the overtaking time. There are two challenges when it comes to the approval of the relations with empirical data. First, The exact value of overtaking time is not known. Second, the exact value of the probability matrix is not accessible, especially after the quarantine policy in each country. In this section, we aim to bring up a novel data analysis method to overcome these challenges and confirm our theoretical achievements with empirical data. We have shown in previous sections that the temporal evolution of infected numbers is known for each node based on the general theory. Also, we know that the most accurate and accessible empirical data is the number of infected people in each node. So If we could rewrite our definition of effective distance in a way that only the number of infected people would be needed, we can achieve a new way to check our claims. Also, using official daily number of patients in different cases, we've estimated each node's overtaking time. We will show that there is a high correlation between effective distance and the estimated overtaking times.  

Consider the number of infectious people versus time. Since the initial value is zero for non-source nodes, one can rewrite the equation for non-source nodes by rescaling time from $t$ into $T=pt$ :

\begin{equation}
I_j=A_j T + B_j T^2
\label{SM-eq:parabola}
\end{equation}

where $I_j$ is the number of patients in the non-source node $j$, $A_j= P_{ij}i_0$, and $B_j= \frac{1}{2} (P{ij}^1+P_{ij}(q_i+q_j-2))$.

With a simple algebra on Eq. \ref{SM-eq:overtaking time12} the effective distance can be rewritten according to $A_j$ and $B_j$ as :

\begin{equation}
D_{ij} = \frac{1}{q_j-\frac{B_j}{A_j}}
\label{SM-eq:effective distance-1}
\end{equation}
It is worth noting that Eq. \ref{SM-eq:effective distance-1} is independent of $i_0$, the initial number of patients, which is challenging to find at the beginning of a pandemic. 

Now, we can calculate $Aj$ and $Bj$  by fitting a parabola to each non-source node patient data. However, there is a likely gap between official and empirical data owing to the fact that it takes several days for governments to identify patients at the beginning. Therefore, to fit, we've used
\begin{equation}
I_j=Q_j+A_j T + B_j T^2
\label{SM-eq:parabola1}
\end{equation}
in which $Q_j$ is the gap.

Moreover, to find $q_j$, it is enough to fit a line to the semi-log plot of patients- time, where it represents an acceptable exponential behavior, resulting from the SIR dynamics. The slop is equal to $q_j$.

The predicted overtaking time is the time when the number of patients is zero. After the coefficients were found, it can be calculated by solving the 

\begin{equation}
0=Q_j+A_j T + B_j T^2
\end{equation}
which leads to 
\begin{equation}
T= \frac{-A_j+\sqrt{A_j^2-4Q_jB_j}}{2A_j}.
\label{SM-eq:the real overtaking time}
\end{equation}

The differences in the regression of lines and number of points in this algorithm come from the accuracy of the raw data and also the mathematical condition ($\frac{P_{ij}^1}{P_{ij}}<2$) that constrain the presence of some nodes in our calculation in different scenarios

It is worth noting that in our formalism the linear relation between Effective distance and overtaking time has y-intercept of zero and the source node has to be in (0,0) naturally. 
But in Figure.\ref{Fig: effective distance-empirical} we observe that y-intercept has a non-zero value, which could be interpreted as a shift in the values of empirical overtaking times.

\section{Using SEIR instead of SIR}
\label{SM-sec:Using SEIR instead of SIR}

If we consider the SEIR dynamic and its equations:

\begin{equation}
    \frac{dE}{dt} = \beta IS - \sigma E
\end{equation}

\begin{equation}
    \frac{dI}{dt} = \sigma E - \gamma I
\end{equation}

By substituting E from the first equation and putting it into the second, we will have :

\begin{equation}
    \frac{dI}{dt} = \beta IS - \gamma I - \frac{dE}{dt}
\end{equation}

In the first stages of the dynamic we can use two assumptions:

First, we can consider the S to be the total population which is N:
\begin{equation}
    S~= N
\end{equation}
Also, we assume that the $\beta I$ is small in comparison to the $\sigma E$ so the first equation would be in the form of :
\begin{equation}
    \frac{dE}{dt} = - \sigma E
\end{equation}

By using this equation we will have:

\begin{equation}
    \frac{dI}{dt} = \beta I N - \gamma I + \sigma E(0)
\end{equation}

Compared to the SIR model, now the solution of our model (by combining the effect of mobility) would be :

\begin{equation}
\Vec{I(t)} = e^{\hat{B}pt} \Vec{I(0)} + \sigma \Vec{E(0)}t
\label{SM-eq:tylorSEIR}
\end{equation}

If we expand this equation, we will have:

\begin{equation}
\Vec{I(t)} = (\hat{I} + \hat{B}pt + ...) \Vec{I(0)} + \sigma \Vec{E(0)}t
\end{equation}

We can summarize the above equation in this form :

\begin{equation}
\Vec{I(t)} = \Vec{I(0)} + (\hat{B}pt \Vec{I(0)} + \sigma \Vec{E(0)})t
\end{equation}
Now by applying the Overtaking condition for node $j$ :
\begin{equation}
\Vec{I(t)}.\vec{A_j} = 0
\end{equation}
we will have :
\begin{equation}
0 = \Vec{I(0)}.\vec{A_j} + (\hat{B}p \Vec{I(0)}.\vec{A_j} + \sigma \Vec{E(0)}.\vec{A_j})t
\end{equation}

So the overtaking time will be:
\begin{equation}
\frac{-\Vec{I(0)}.\vec{A_j}}{(\hat{B} \Vec{I(0)}.\vec{A_j} + \frac{\sigma}{p} \Vec{E(0)}.\vec{A_j})}= pt_o
\end{equation}

By using Taylor expansion :

\begin{equation}
\frac{-\Vec{I(0)}.\vec{A_j}}{(\hat{B} \Vec{I(0)}.\vec{A_j} + \frac{\sigma}{p} \Vec{E(0)}.\vec{A_j})}= pt_o
\end{equation}

\section{Data}\label{SM-sec:Data}

In this study, we primarily utilized two different types of data: A) Snapshots of active infected cases in each subpopulation during the linear phase, some of which are visualized in Fig. \ref{fig: Where}, panels $A$ and $A^{\prime}$, and B) Coarse-grained representations of inter-population mobility, illustrated in Fig. \ref{fig: Where}, panels $B$ and $B^{\prime}$. Both data types were used for Figs. \ref{fig: Where} and \ref{fig: when}, while only mobility data were used for Fig. \ref{fig: Effective distance}, and only infected cases were used for Fig. \ref{Fig: effective distance-empirical}, see section \ref{SM-sec:How to estimate Effective Distance without mobility data} for more details.

We obtained the number of infected people during the COVID-19 pandemic for different provinces of Iran from official reports by the Ministry of Health and Medical Education of Iran, which are available in Persian at request, and from The COVID Tracking Project at The Atlantic \cite{COVIDTrackingProject} for different states of the US. Data regarding the H1N1 pandemic was downloaded from "www.who.int". Abbreviations for Iranian provinces and American states are listed in Sec. \ref{SM-sec:ListAbbreviation}.

The daily mobility data, which encompasses all forms of transportation for Iran (provided by the Basir company) and the USA \cite{kang2020multiscale}, is averaged from March 1st to 3rd, 2020, for Iran, and from January to April 2020 for the USA.

\subsection{Abbreviation}
\label{SM-sec:ListAbbreviation}
\begin{table}[ht]
\begin{center}

\begin{tabular}{ |p{2.4cm}|p{1cm}|p{2.4cm}|p{1cm}|p{2.4cm}|p{1cm}|p{2.4cm}|p{1cm}|}
 \hline

 State  & Abv. & State & Abv.& State & Abv.& State & Abv.\\
 \hline
Alabama & AL & Alaska & AK & Arizona & AZ & Arkansas & AR\\
California & CA & Colorado & CO & Connecticut & CT & Delaware & DE\\
Columbia & DC & Florida & FL & Georgia & GA & Idaho & ID \\
Kentucky & KY & Louisiana & LA & Maine & ME & Maryland & MD\\
Illinois & IL & Indiana & IN & Iowa & IA & Kansas & KS\\
Kentucky & KY & Louisiana & LA & Maine & ME & Maryland & MD \\ Massachusetts & MA & Michigan & MI & Minnesota & MN & Mississippi & MS \\ Missouri & MO & Montana & MT & Nebraska & NE & Nevada & NV \\ NewHampshire & NH & New Jersey & NJ & New Mexico & NM & New York & NY \\ North Carolina & NC & North Dakota & ND & Ohio & OH & Oklahoma & OK \\ Oregon & OR & Pennsylvania & PA & Rhode Island & RI & South Carolina & SC \\ South Dakota & SD & Tennessee & TN & Texas & TX & Utah & UT \\ Vermont & VT & Virginia & VA & Washington & WA & West Virginia & WV \\ Wisconsin & WI & Wyoming & WY & -- & -- & -- & --\\
 \hline
\end{tabular}
\caption{Abbreviation of the US states.}
\label{SM-table: the US states}
\end{center}
\end{table}

\begin{table}
\begin{center}
\begin{tabular}{ |p{4cm}|p{1cm}|p{4cm}|p{1cm}|}
 \hline
 Province  & Abv. & Province & Abv.\\
 \hline
Ardabil & AR & Sistan-Baluchestan & SB \\Ilam & IL & Golestan & GL \\Khorasan-North & KS & Gilan & GI \\Kurdistan-South & KJ & Yazd & YZ \\Kermanshah & KM & Zanjan & ZN \\Kordestan & KD & Kohgiluyeh-Boyer-Ahmad & KB \\Bushehr & BU & ChaharMahaal-Bakhtiari & CB \\Khorasan-Razavi & KR & Mazandaran & MZ \\Khuzestan & KZ & Hormozgan & HR \\Kerman & KN & Hamadan & HM \\Lorestan & LR & Azerbaijan-West & AG \\Qom & QM & Azerbaijan-East & AS \\Semnan & SM & Fars & FR \\Isfahan & ES & Qazvin & QZ\\Markazi & MK & Alborz & AL\\Tehran & TH & & \\
 \hline
\end{tabular}
\end{center}
\caption{Abbreviation of provinces of Iran.}
\label{SM-table: Iran provinces}
\end{table}
\newpage

\section{Sensitivity Analysis}
\label{SM-sec:Sensitivity Analysis}
These plots represent the sensitivity analysis in which $\gamma$ has been changed from $\frac{1}{13}$ to $\frac{1}{20}$ and $R_0$ which is the basic reproductive number has been changed from 2.5 to 4.5. By using the day 5 from the official start of the pandemic we can say the 5 maximum provinces in the node power list is Qom, Tehran, Gilan, Markazi and Alborz. These provinces stay sorted in this format for all the values of $R_0$ and $\gamma$. As the $R_0$ increases the value of error increases too. As $\gamma$ decreases the value of error decreases. But by looking at 10 days we can see that the place of Tehran and Qom has been swaped. Alborz came to the third place and got the previous position of Gilan. But pay attention that all these names remain reserved in the 10 days. But in 20 days Semnan appears in the list.
\begin{figure}[ht]
\centering
\includegraphics[width=\textwidth]{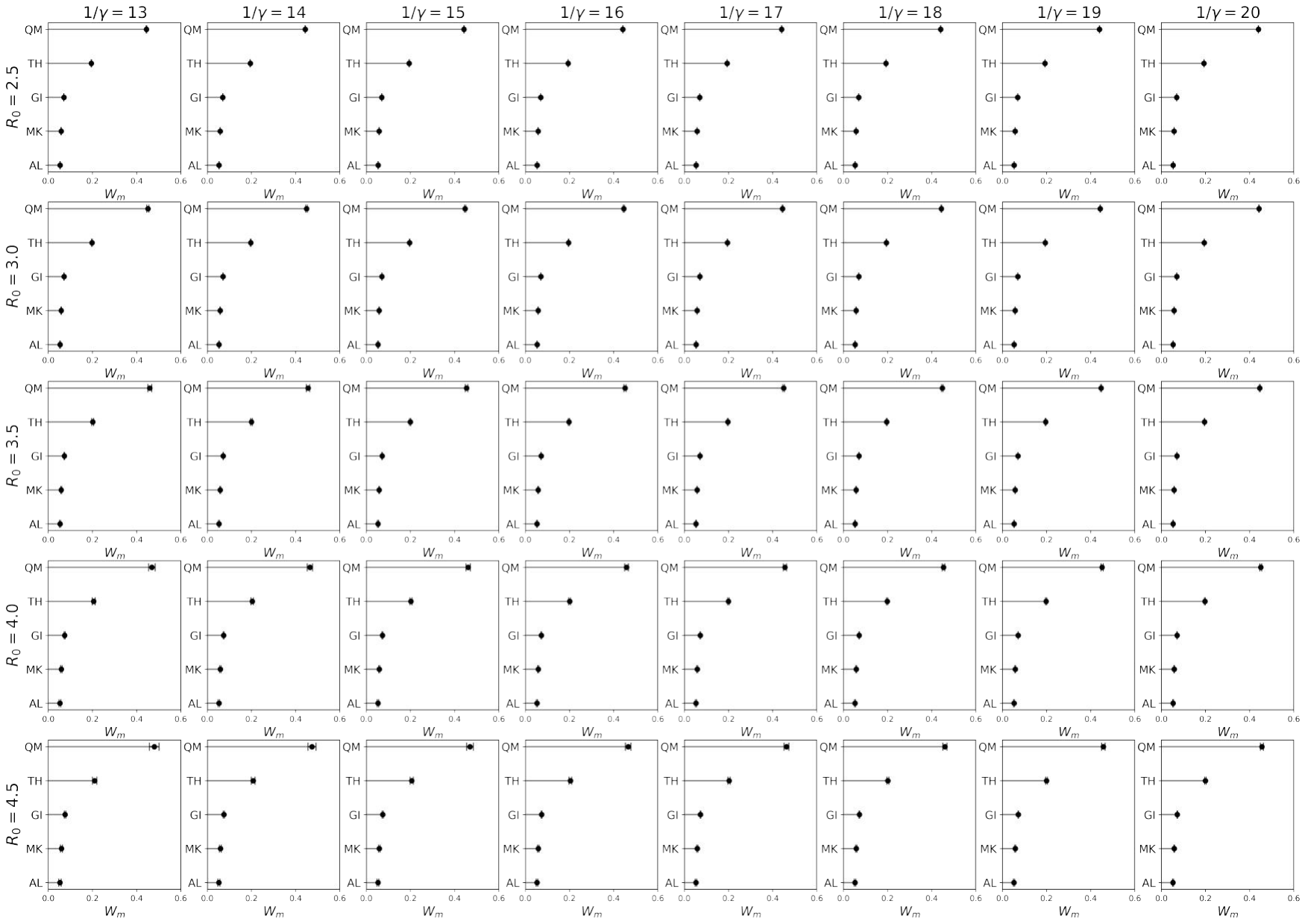}
\caption{Calculated $W_m$ for different provinces of Iran, and for different values of $R_0$ and $\gamma$, using the number of infected people in the fifth day of the COVID-19 pandemic in Iran.}
\end{figure}
\begin{figure}[ht]
\centering
\includegraphics[width=\textwidth]{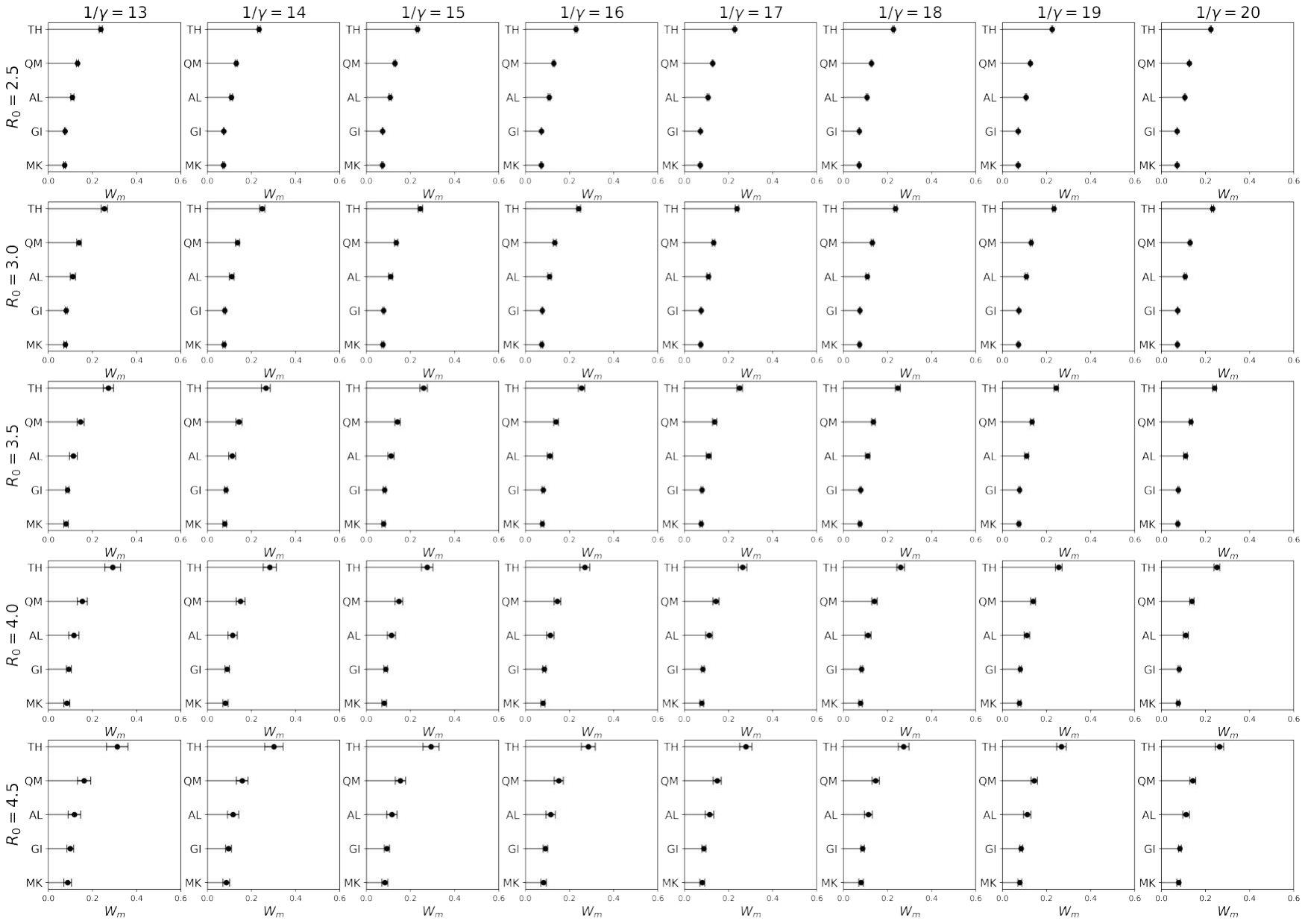}
\caption{Calculated $W_m$ for different provinces of Iran, and for different values of $R_0$ and $\gamma$, using the number of infected people in the tenth day of the COVID-19 pandemic in Iran.}
\end{figure}
\begin{figure}[ht]
\centering
\includegraphics[width=\textwidth]{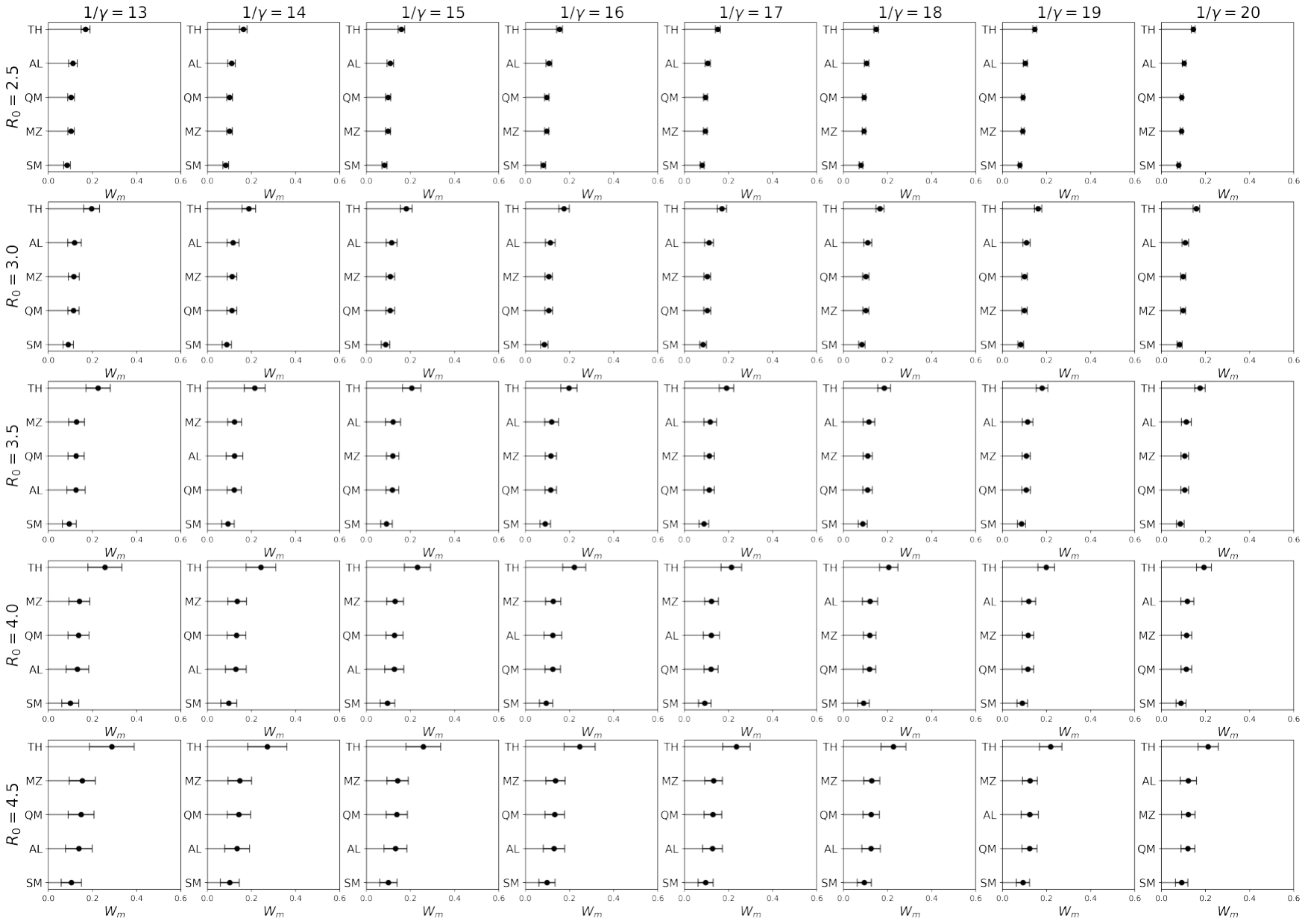}
\caption{Calculated $W_m$ for different provinces of Iran, and for different values of $R_0$ and $\gamma$, using the number of infected people in the fifteenth day of the COVID-19 pandemic in Iran.}
\end{figure}
\begin{figure}[ht]
\centering
\includegraphics[width=\textwidth]{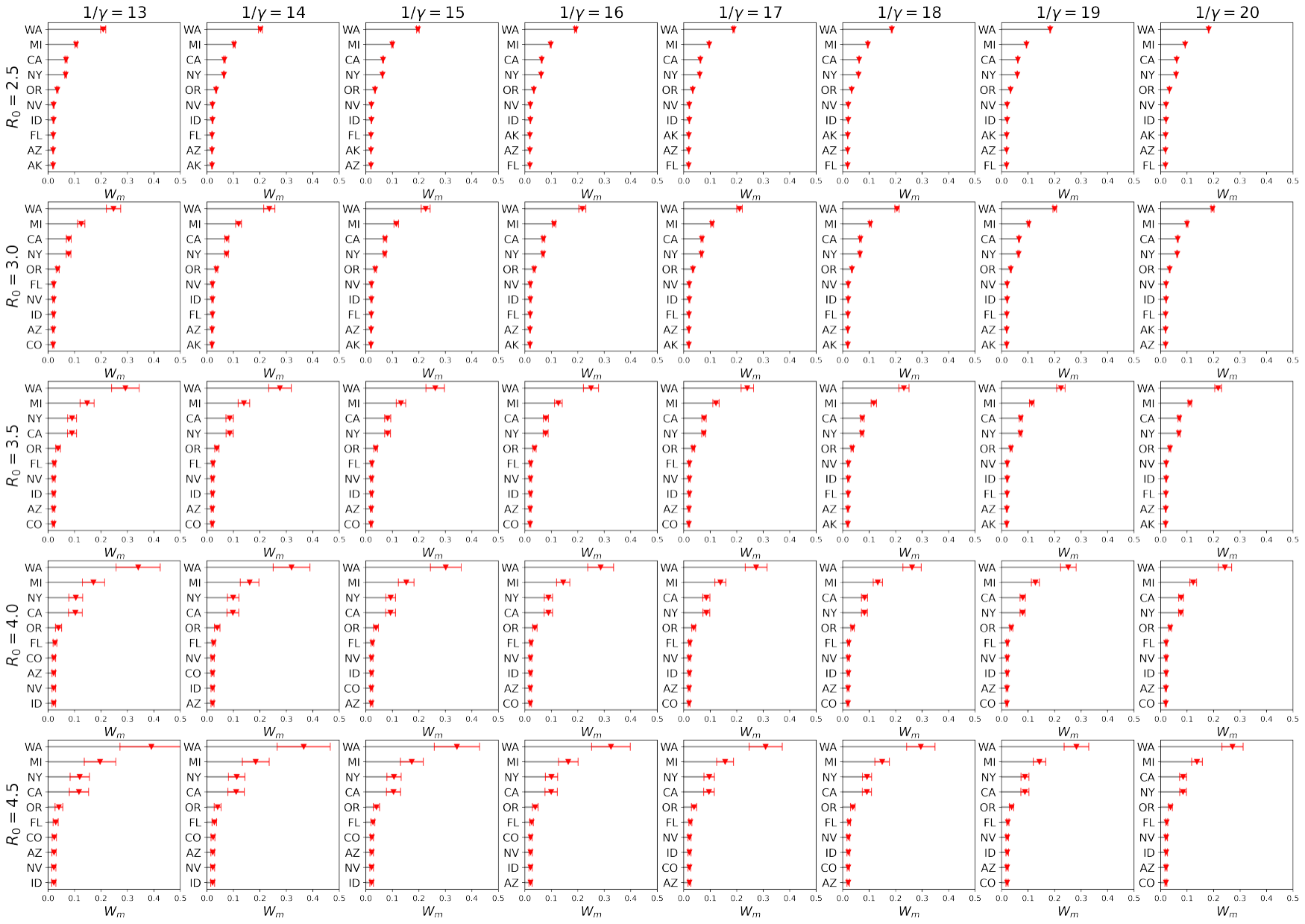}
\caption{Calculated $W_m$ for different states of the US, and for different values of $R_0$ and $\gamma$, using the number of infected individuals in day 45 of the COVID-19 pandemic in the US.}
\end{figure}
\begin{figure}[ht]
\centering
\includegraphics[width=\textwidth]{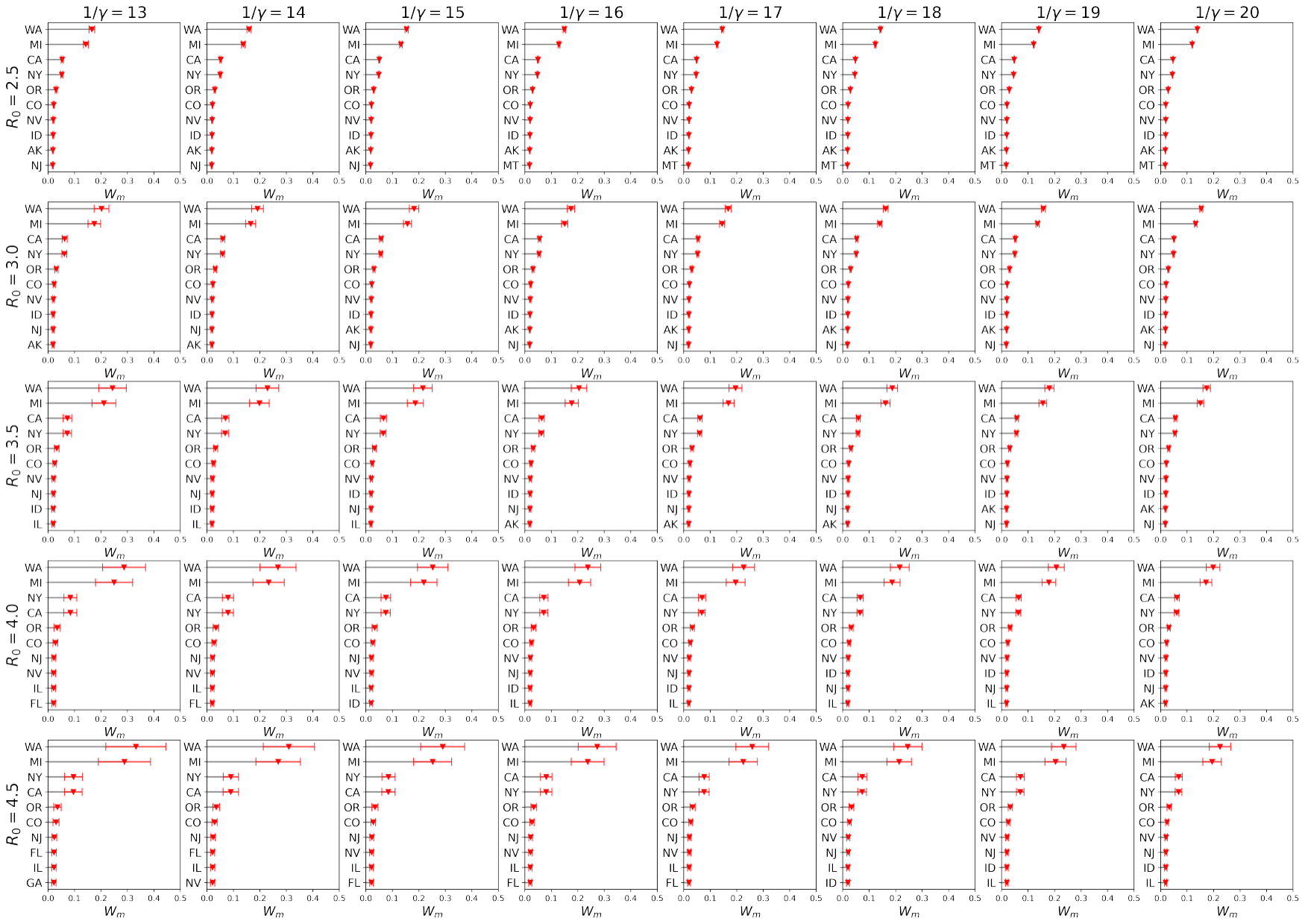}
\caption{Calculated $W_m$ for different states of the US, and for different values of $R_0$ and $\gamma$, using the number of infected individuals in day 50 of the COVID-19 pandemic in the US.}
\end{figure}
\begin{figure}[ht]
\centering
\includegraphics[width=\textwidth]{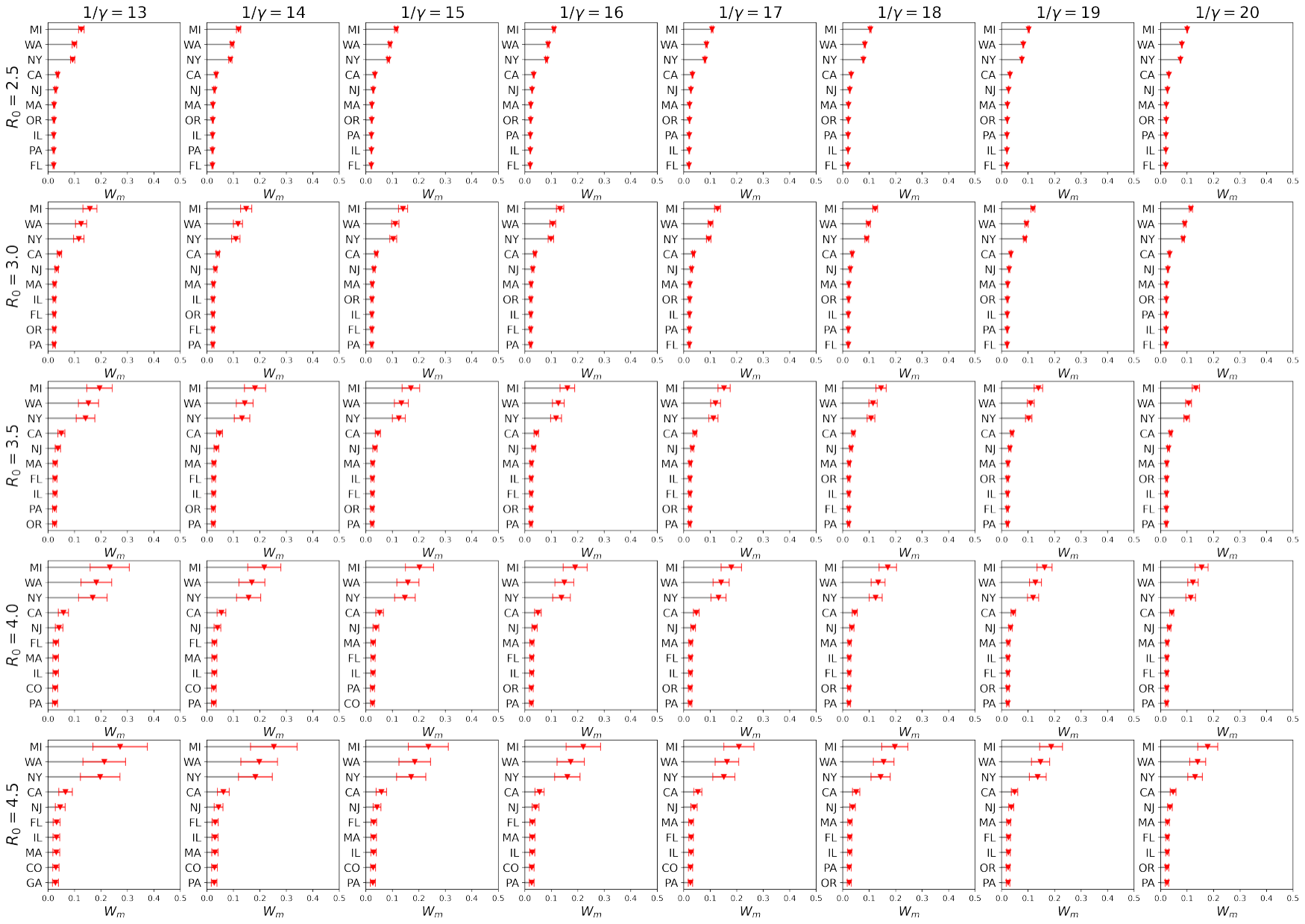}
\caption{Calculated $W_m$ for different states of the US, and for different values of $R_0$ and $\gamma$, using the number of infected individuals in day 55 of the COVID-19 pandemic in the US.}
\end{figure}

\end{document}